\documentclass[
reprint,
superscriptaddress,
nofootinbib,
floatfix,
amsmath,
amssymb,
aps,
prd
]{revtex4-2}
\usepackage{graphicx}
\usepackage{dcolumn}
\usepackage{bm}
\usepackage[caption=false]{subfig}
\usepackage{booktabs}
\usepackage{longtable}
\usepackage{multirow}
\usepackage{url}

\begin{document}

\title{Latent-Space Gaussian Processes for Dark-Energy Reconstruction from Observational \(H(z)\) Data}

\author{Jia-Yan Jiang}
\author{Wei Hong}
\author{Tong-Jie Zhang}
\email{tjzhang@bnu.edu.cn}

\affiliation{Institute for Frontiers in Astronomy and Astrophysics, Beijing Normal University, Beijing 102206, People's Republic of China}
\affiliation{School of Physics and Astronomy, Beijing Normal University, Beijing 100875, People's Republic of China}

\date{\today}

\begin{abstract}
Using the 37-point cosmic-chronometer subset of observational Hubble parameter (OHD) data, we develop a Bayesian Gaussian-process framework to reconstruct the normalized dark-energy density \(f(z)\) and equation of state \(w(z)\), focusing on how the choice of latent space affects the inference. We compare a Gaussian-process prior placed directly on \(f(z)\) with the conventional latent-\(H\) formulation, and also test a log-\(f\) branch that enforces \(f(z)>0\). We further analyze OHD-like mock data generated from fiducial \(\Lambda\)CDM and mildly evolving \(w_0w_a\) models, using both the observed redshift distribution and a higher-quality high-redshift setup. For real OHD, leave-one-out cross-validation shows no strong predictive preference between latent-\(f\) and latent-\(H\) reconstructions. The inferred \(f(z)\), \(w(z)\), and \(Om(z)\) remain consistent with \(\Lambda\)CDM across the tested external priors, while apparent \(Om(z)\) trends are prior sensitive and not robust evidence for dark-energy evolution. Residual differences between the two latent constructions are small, sign mixed, prior dependent, and mainly confined to the weakly constrained high-redshift tail. We therefore interpret the real-data results primarily as a methodological assessment. In mock tests, the framework responds to injected mild evolution in the reconstructed dark-energy quantities and \(Om(z)\), with detectability depending on method and data coverage. Improved high-redshift OHD reduces the discrepancy between latent constructions and makes the \(Om(z)\) response more consistently detectable. The latent-\(f\) approach is therefore a viable alternative to latent-\(H\), while current constraints are limited mainly by sparse high-redshift OHD and dependence on external priors.

\end{abstract}
\maketitle

\section{Introduction}

The physical origin of cosmic acceleration remains one of the central problems in modern cosmology.
Within the standard $\Lambda$CDM framework, the accelerated expansion is attributed to a cosmological constant with equation of state $w=-1$ \cite{Weinberg1989,carroll2001cosmological,bull2016beyond,efstathiou2025challenges}. Despite its remarkable success in describing cosmological observations, the cosmological constant raises long-standing theoretical difficulties \cite{martin2012everything,rugh2002quantum,scali2025cosmological}. These considerations motivate robust reconstruction methods for testing whether the dark-energy density evolves with cosmic time.

A common strategy is to reconstruct the dark-energy equation of state $w(z)$ from observational Hubble parameter data (OHD).
Parametric descriptions, such as the widely used $w_0$--$w_a$ model \cite{chevallier2001accelerating,linder2003exploring}, provide a convenient framework but restrict the allowed functional forms of dark-energy evolution.
Nonparametric methods offer a more flexible alternative.
Among them, Gaussian-process (GP) regression has become a widely used flexible nonparametric tool for reconstructing cosmological functions from data \cite{holsclaw2010nonparametric,holsclaw2011nonparametric,2012JCAP...06..036S,shafieloo2012gaussian,zhao2017dynamical,li2025reconstructing,jiang2025optimizing,escamilla2023model,velazquez2024non,yang2025gaussian,zhang2026reconstruction,johnson2025kernel}.
In a GP reconstruction, the unknown function of interest is modeled as a latent Gaussian process, while the covariance kernel encodes assumptions about smoothness and correlation structure \cite{titsias2010bayesian,williams2006gaussian,li2016review,alvarez2009latent,nguyen2009model}.

In most GP reconstructions, the Hubble parameter $H(z)$ is treated as the latent quantity, and other cosmological functions are derived from it.
In particular, the dark-energy density function $f(z)$ and the equation-of-state parameter $w(z)$ are obtained through nonlinear transformations of the reconstructed $H(z)$ \cite{yu2013nonparametric,2012JCAP...06..036S,velazquez2024non,perenon2022measuring,wang2017improved,zhang2018gaussian}.
Such a pipeline introduces an additional layer of uncertainty propagation because the derived quantities depend on nonlinear combinations of $H(z)$ and its derivative. The nonlinear transformations from \(H(z)\) to \(f(z)\) and \(w(z)\), especially those involving derivatives, can amplify numerical fluctuations, further motivating alternative latent choices \cite{johnson2025kernel,jesus2024hubble,dutta2020beyond}.

These considerations motivate exploring alternative choices of latent variables.
In this work, we reconstruct the normalized dark-energy density $f(z)$ directly using a latent Gaussian process (Method~A) and compare this approach with the conventional $H(z)$-latent reconstruction (Method~B).
Because $f(z)$ directly characterizes dark-energy evolution, this latent choice may reduce one stage of nonlinear propagation and provides a controlled way to assess the impact of latent representation on the inferred dark-energy reconstruction \cite{linder2004reconstructing,lodha2025extended}.

In the real-data analysis, we restrict attention to the cosmic-chronometer (CC) subset of OHD, namely \(H(z)\) measurements inferred from differential ages of passively evolving galaxies, which have been widely used as an independent cosmological probe \cite{2010AdAst2010E..81Z,2011ApJ...730...74M,jiao2023new}.

The equation of state depends on the derivative-to-amplitude ratio \(f'(z)/f(z)\). Consequently, posterior draws in which \(f(z)\) approaches zero can produce large excursions in the inferred \(w(z)\). This numerical sensitivity is not specific to either latent construction, since both the latent-\(H\) and latent-\(f\) pipelines ultimately evaluate the same ratio when deriving \(w(z)\) \cite{sahni2003braneworld,sahni2008two,tsujikawa2008constraints,bauer2010dynamically,sahni2014model,gomez2015background,dai2018reconstruction,akarsu2019screening,akarsu2020graduated,di2022sign,moshafi2022multiple,ozulker2022dark,ong2023effective,adil2024omnipotent}.

In practice, this sensitivity is mitigated by our fully Bayesian GP treatment, in which latent functions and hyperparameters are jointly sampled and posterior summaries are reported instead of relying on point estimates \cite{williams2006gaussian,2012JCAP...06..036S}. This treatment reduces the influence of isolated pole-crossing draws, although it does not remove the underlying derivative-to-amplitude sensitivity. 

The derivative-to-amplitude sensitivity also motivates the constrained log-\(f\) branch: since \(w(z)\) depends on \({\rm d}\ln f/{\rm d}z\), a GP prior on \(\ln f(z)\) acts directly on the latent quantity controlling \(w(z)\) while enforcing \(f(z)>0\) by construction. We use this branch as a constrained sensitivity test for positivity and suppression of zero-crossings in \(f(z)\), rather than as the primary reconstruction.

Reconstruction strategies are compared using Pareto-smoothed importance-sampling leave-one-out cross-validation (PSIS-LOO) \cite{vehtari2016bayesian,vehtari2017practical,magnusson2019bayesian,magnusson2020leave,vehtari2024pareto}, a computationally efficient approximation to leave-one-out predictive assessment. This allows us to compare latent constructions, covariance kernels, and prior choices within a common predictive framework. Cross-method differences are quantified via pointwise effect-size and posterior-direction diagnostics together with interval ROPE summaries, while sensitivity to external priors \((H_0,\Omega_{m0},\Omega_{k0})\) is examined. Consistency with \(\Lambda\)CDM is assessed using both \(f(z)\) and the \(Om(z)\) diagnostic \cite{sahni2008two,sahni2014model,qi2018parameterized,myrzakulov2023new}.

Finally, we complement the real-data analysis with OHD-like mock datasets generated from both a fiducial \(\Lambda\)CDM truth model and a mildly evolving \(w_0w_a\) truth model. By comparing the observed redshift distribution with a modified setup in which the high-redshift tail is sampled more densely and precisely, we use these mocks to test whether the residual A/B ambiguity is primarily data limited, especially in the sparse \(z\gtrsim1.5\) tail, and whether the same framework and the \(Om(z)\) diagnostic can track the injected evolution.

In this work, we examine whether the latent-\(f\) strategy is practically feasible with the current CC subset of OHD, whether it differs predictively from the conventional latent-\(H\) strategy under PSIS-LOO, and how strongly the reconstructed dark-energy quantities depend on external priors and on the constrained log-\(f\) branch. We also use OHD-like mock tests, including cases with denser and more precise sampling in the high-redshift tail, to investigate the origin of the present high-redshift ambiguity and the detectability of mild evolution in \(Om(z)\). Our goal is not to claim a real-data detection of dynamical dark energy from the current CC subset of OHD. Instead, we use this analysis to examine how methodological choices and present data limitations jointly shape the robustness, predictive comparison, and physical interpretation of nonparametric dark-energy reconstruction. For the current CC subset of OHD, this means that any sensitivity to reconstruction strategy is concentrated mainly in the weakly constrained high-redshift tail, while the dominant limitation comes from the sparsity of the high-\(z\) data and their residual dependence on external priors.

The remainder of the paper is organized as follows. Section~\ref{sec:theory} summarizes the cosmological framework. Sections~\ref{sec:data}, \ref{sec:priors}, and \ref{sec:gp} describe the data, external priors, and Bayesian GP reconstruction methods. Section~\ref{sec:results} presents the real-data reconstructions, predictive model comparison, and mock-data tests. Section~\ref{sec:conclusion} summarizes our conclusions.

\section{Cosmological framework}
\label{sec:theory}
We work in the Friedmann--Lema\^itre--Robertson--Walker framework with matter, radiation, spatial curvature, and dark energy \cite{weinberg2008cosmology,dodelson2020modern}. The expansion history is
\begin{equation}
\begin{aligned}
H^2(z)=H_0^2\!\Bigl[&
\Omega_{m0}(1+z)^3+\Omega_{r0}(1+z)^4\\
&+\Omega_{k0}(1+z)^2+\Omega_{\mathrm{DE}0} f(z)
\Bigr],
\end{aligned}
\label{eq:flrw}
\end{equation}

where $H_0$ is the Hubble constant, and $\Omega_{m0}$, $\Omega_{r0}$, $\Omega_{k0}$, and $\Omega_{\mathrm{DE}0}$ are the present-day density parameters of matter, radiation, curvature, and dark energy, respectively. These satisfy
\begin{equation}
\Omega_{m0}+\Omega_{r0}+\Omega_{k0}+\Omega_{\mathrm{DE}0}=1.
\end{equation}
The normalized dark-energy density is defined by
\begin{equation}
f(z)=\exp\!\left[3\int_0^z \frac{1+w(z')}{1+z'}\,\mathrm{d}z'\right].
\end{equation}

In the late-time regime considered here, radiation is negligible, while a small spatial-curvature component is allowed \cite{cosmologyscott,2020MNRAS.496L..91E}. Neglecting radiation gives \(\Omega_{\mathrm{DE}0}=1-\Omega_{m0}-\Omega_{k0}\), and the late-time Friedmann equation can be solved for \(f(z)\) as

\begin{equation}
f(z)=\frac{H^2(z)/H_0^2-\Omega_{m0}(1+z)^3-\Omega_{k0}(1+z)^2}{1-\Omega_{m0}-\Omega_{k0}}.
\label{eq:f_of_H}
\end{equation}
By definition, the normalized dark-energy density satisfies \(f(0)=1\). For flat $\Lambda$CDM, $w(z)=-1$ and therefore $f(z)=1$ at all redshifts. Defining $f'(z)\equiv \mathrm{d}f/\mathrm{d}z$, the dark-energy equation of state may be written as
\begin{equation}
w(z)=-1+\frac{1+z}{3}\frac{f'(z)}{f(z)}.
\label{eq:w_of_f}
\end{equation}
Using Eq.~\eqref{eq:f_of_H}, we obtain
\begin{equation}
f'(z)=
\frac{
\frac{2H(z)}{H_0^2}\frac{\mathrm{d}H}{\mathrm{d}z}
-3\Omega_{m0}(1+z)^2
-2\Omega_{k0}(1+z)
}{
1-\Omega_{m0}-\Omega_{k0}
}.
\label{eq:fprime_from_H}
\end{equation}
Equation~\eqref{eq:w_of_f} is singular when $f(z)=0$ \cite{sahni2003braneworld,sahni2008two,sahni2014model}, equivalently (for $1-\Omega_{m0}-\Omega_{k0}\neq 0$),
\begin{equation}
H^2(z)=H_0^2\!\left[\Omega_{m0}(1+z)^3+\Omega_{k0}(1+z)^2\right].
\end{equation}
This singularity is a mathematical feature of the \(w(z)\) parametrization, arising from the vanishing denominator in Eq.~\eqref{eq:w_of_f}. It is therefore natural to regard $f(z)$ as the primary reconstructed quantity and to treat $w(z)$ as a derived quantity. Since $w(z)$ involves both a derivative and a ratio, it is generically more sensitive to reconstruction noise than $f(z)$, especially when the reconstructed $f(z)$ approaches zero. Apparent structure in $w(z)$ should therefore be interpreted more cautiously than the corresponding behavior of $f(z)$.

The function \(f(z)\) also provides a convenient route to consistency tests of \(\Lambda\)CDM. A widely used example is the \(Om(z)\) diagnostic, here written in a form that allows for spatial curvature \cite{Om2008sahni},
\begin{equation}
Om(z)=\frac{H^2(z)/H_0^2-1-\Omega_{k0}\!\left[(1+z)^2-1\right]}{(1+z)^3-1}.
\end{equation}
Using Eq.~\eqref{eq:f_of_H}, this may be rewritten as
\begin{equation}
Om(z)=\Omega_{m0}+(1-\Omega_{m0}-\Omega_{k0})
\frac{f(z)-1}{(1+z)^3-1}.
\end{equation}
Under this definition, which allows for spatial curvature, $\Lambda$CDM gives $Om(z)=\Omega_{m0}$ as a constant. For two ordered redshifts \(z_i<z_j\), we therefore also consider the two-point statistic
\begin{equation}
\Delta Om_{ij}=Om(z_i)-Om(z_j),
\label{eq:delta_om}
\end{equation}
which satisfies $\Delta Om_{ij}=0$ for a cosmological constant \cite{Om2008sahni}. For $z_1<z_2$, the standard sign test is summarized in Table~\ref{tab:om_sign_rule}. In the present nonparametric setting, we use it as a qualitative diagnostic rather than as a proof of constant-\(w\) behavior.

\begin{table}[!htbp]
\centering
\caption{Qualitative sign interpretation of the two-point \(Om\) diagnostic under the convention \(\Delta Om_{12}=Om(z_1)-Om(z_2)\) for \(z_1<z_2\).}
\label{tab:om_sign_rule}
\begin{tabular}{ll}
\toprule
Criterion on $\Delta Om_{12}$ & Interpretation \\
\midrule
$\Delta Om_{12}=0$ & $\Lambda$CDM-like ($w=-1$) \\
$\Delta Om_{12}>0$ & Quintessence-like ($w>-1$) \\
$\Delta Om_{12}<0$ & Phantom-like ($w<-1$) \\
\bottomrule
\end{tabular}
\end{table}

One may place the GP prior directly on \(f(z)\) and map it forward to the
observed \(H(z)\) likelihood, or instead place the GP prior on \(H(z)\) and
subsequently map the reconstructed \(H(z)\) to \(f(z)\) through
Eq.~\eqref{eq:f_of_H}. In both cases, $w(z)$ is treated as a derived quantity through Eq.~\eqref{eq:w_of_f}. These two strategies are developed explicitly in Sec.~\ref{sec:gp} and, for notational simplicity, will be referred to below as Method~A and Method~B, respectively.

\section{Datasets and methodology}
\subsection{Data}
\label{sec:data}
We use OHD obtained via the CC method, inferred from differential-age estimates of passively evolving galaxies \cite{2002ApJ...573...37J,Moresco_2022}. The dataset adopted in this work contains $n=37$ measurements spanning $0.07 \le z \le 1.965$ in Table~\ref{tab:ccdata}. One entry at \(z=0.09\), reported in the original literature as
\(H_0=69\pm12\ \mathrm{km\,s^{-1}\,Mpc^{-1}}\), is reformulated here as a
redshift-dependent Hubble measurement by applying the standard
\(\Lambda\)CDM evolution relation adopted in that work,
\(H(z)=H_0E(z)\), using the same parameter values. This gives
\(H(z=0.09)=70.7\pm12.3\ \mathrm{km\,s^{-1}\,Mpc^{-1}}\).
We mark this converted point explicitly in Table~\ref{tab:ccdata} and include it in the final sample.

We treat the measurements as statistically independent, so that the covariance matrix is taken to be diagonal. We note, however, that nontrivial covariance information has been derived for part of the CC sample, particularly for D4000-based measurements, and that stellar-population-synthesis systematics may induce additional inter-point correlations \cite{moresco_cccovariance,2020ApJ...898...82M}. Since the present 37-point compilation is assembled from different CC analyses and surveys, with full covariance information unavailable in a homogeneous form for the entire sample, we adopt the diagonal approximation in the present analysis. This choice is also consistent with recent GP-based studies showing that, for current limited and heterogeneous CC compilations, diagonal treatments can perform competitively relative to sparse full-covariance implementations \cite{jiang2025optimizing}.

\begin{table}[t]
\centering
\begin{tabular}{llllllll}
\toprule
$z$ & $H(z)$ & $\sigma_{H(z)}$ & Ref. & $z$ & $H(z)$ & $\sigma_{H(z)}$ & Ref. \\
\midrule
0.07 & 69 & 19.6 & \cite{zhang2014four} & 0.48 & 97.0 & 62.0 & \cite{stern2010cosmic} \\
0.09 & 70.7 & 12.3 & \cite{jimenez2003constraints}$\ast$ & 0.5 & 72.1 & 34.68 & \cite{loubser2025independent} \\
0.12 & 68.6 & 26.2 & \cite{zhang2014four} & 0.593 & 104.0 & 13.0 & \cite{moresco2012improved} \\
0.17 & 83.0 & 8.0 & \cite{simon2005constraints} & 0.67 & 119.45 & 17.82 & \cite{loubser2025measuring} \\
0.179 & 75.0 & 4.0 & \cite{moresco2012improved} & 0.68 & 92.0 & 8.0 & \cite{moresco2012improved} \\
0.199 & 75.0 & 5.0 & \cite{moresco2012improved} & 0.781 & 105.0 & 12.0 & \cite{moresco2012improved} \\
0.2 & 72.9 & 29.6 & \cite{zhang2014four} & 0.8 & 113.1 & 25.22 & \cite{jiao2023new} \\
0.27 & 77.0 & 14.0 & \cite{simon2005constraints} & 0.83 & 108.28 & 18.13 & \cite{loubser2025measuring} \\
0.28 & 88.8 & 36.6 & \cite{zhang2014four} & 0.8754 & 125.0 & 17.0 & \cite{moresco2012improved} \\
0.352 & 83.0 & 14.0 & \cite{moresco2012improved} & 0.88 & 90.0 & 40.0 & \cite{stern2010cosmic} \\
0.38 & 83.0 & 13.5 & \cite{moresco20166} & 0.9 & 117.0 & 23.0 & \cite{simon2005constraints} \\
0.4 & 95.0 & 17.0 & \cite{simon2005constraints} & 1.037 & 154.0 & 20.0 & \cite{moresco2012improved} \\
0.4004 & 77.0 & 10.2 & \cite{moresco20166} & 1.26 & 135.0 & 65.0 & \cite{tomasetti2023new} \\
0.425 & 87.1 & 11.2 & \cite{moresco20166} & 1.3 & 168.0 & 17.0 & \cite{simon2005constraints} \\
0.445 & 92.8 & 12.9 & \cite{moresco20166} & 1.363 & 160.0 & 33.6 & \cite{moresco2015raising} \\
0.46 & 88.48 & 12.33 & \cite{loubser2025measuring} & 1.43 & 177.0 & 18.0 & \cite{simon2005constraints} \\
0.47 & 89.0 & 49.6 & \cite{ratsimbazafy2017age} & 1.53 & 140.0 & 14.0 & \cite{simon2005constraints} \\
0.4783 & 80.9 & 9.0 & \cite{moresco20166} & 1.75 & 202.0 & 40.0 & \cite{simon2005constraints} \\
& & & & 1.965 & 186.5 & 50.4 & \cite{moresco2015raising} \\
\bottomrule
\end{tabular}
\caption{OHD used in this work. The \(H(z)\) values and uncertainties are in \(\mathrm{km\,s^{-1}\,Mpc^{-1}}\). ``Ref.'' denotes the literature source of each measurement. An asterisk (\(\ast\)) denotes a value converted from an \(H_0\) estimate under the standard redshift-evolution model.}
\label{tab:ccdata}
\end{table}

\subsection{External cosmological inputs}
\label{sec:priors}

The mapping from \(H(z)\) to \(f(z)\) and \(w(z)\) requires external inputs for \(H_0\), \(\Omega_{m0}\), and \(\Omega_{k0}\). We therefore consider three external prior sets, denoted Planck2018, DESI--BBN, and SH0ES--Pantheon, which differ in the adopted priors for \(H_0\) and \(\Omega_{m0}\). For \(\Omega_{k0}\), we use the same Gaussian prior in all three cases, taken from the Planck2018 constraint. This choice keeps the curvature prior fixed so that the three cases mainly probe the impact of different \(H_0\) and \(\Omega_{m0}\) prior sets. The Planck2018 prior set adopts the \(\Lambda\)CDM constraints from the Planck CMB analysis \cite{planck2018}. The DESI--BBN prior set is based on DESI BAO measurements combined with Big Bang nucleosynthesis information \cite{desi2025}. The SH0ES--Pantheon prior set combines the local distance-ladder determination of \(H_0\) from SH0ES with the Pantheon+ constraint on \(\Omega_{m0}\) from Type Ia supernovae \cite{riess2022comprehensive,brout2022pantheonplus}.

In the main analysis, we perform the reconstruction separately for each prior set and place Gaussian priors on \((H_0,\Omega_{m0},\Omega_{k0})\), with means and \(1\sigma\) widths listed in Table~\ref{tab:external_priors}.

\begin{table}[!htbp]
\centering
\caption{External cosmological inputs adopted in this work. \(H_0\) is given in \(\mathrm{km\,s^{-1}\,Mpc^{-1}}\).}
\label{tab:external_priors}
\begin{tabular}{@{}lccc@{}}
\toprule
Prior & $H_0$ & $\Omega_{m0}$ & $\Omega_{k0}$ \\
\midrule
Planck2018 & $67.4\pm0.5$ & $0.315\pm0.007$ & $0.001\pm0.002$ \\
DESI--BBN & $67.98\pm0.75$ & $0.295\pm0.015$ & $0.001\pm0.002$ \\
SH0ES--Pantheon & $73.04\pm1.04$ & $0.334\pm0.018$ & $0.001\pm0.002$ \\
\bottomrule
\end{tabular}
\end{table}

\subsection{Gaussian-process reconstruction}
\label{sec:gp}

We denote the OHD by
\begin{equation}
\mathcal D=\{(z_i,H_i,\sigma_{H,i})\}_{i=1}^n.
\end{equation}
For each reconstruction model $\mathcal M$, specified by a choice of latent construction and covariance kernel, we introduce a latent function $g(z)$.

Let $\mathbf z=(z_1,\dots,z_n)^\top$ be the observed redshifts and let $\mathbf g=(g_1,\dots,g_n)^\top$, with $g_i=g(z_i)$, denote the latent-function values at those locations. We place a zero-mean Gaussian-process prior on $\mathbf g$,
\begin{equation}
\mathbf g \mid \eta,\mathcal M
\sim
\mathcal N\!\bigl(\mathbf 0,\mathbf K_\eta(\mathbf z,\mathbf z)\bigr),
\end{equation}
where $\mathbf K_\eta$ is the covariance matrix induced by the kernel $k_\eta$, with entries $[\mathbf K_\eta]_{ij}=k_\eta(z_i,z_j)$. The hyperparameters $\eta=(\ell,\sigma_g)$ control the correlation length scale $\ell$ and overall amplitude $\sigma_g$. The zero mean function is used as a conventional reference choice rather than as a physical assumption about the reconstructed quantity. In the present nonparametric setting, the functional flexibility is controlled primarily by the covariance kernel and its hyperparameters, which are inferred jointly with the latent function. We consider six covariance kernels: RBF, Mat\'ern $3/2$, Mat\'ern $5/2$, Mat\'ern $7/2$, Mat\'ern $9/2$, and Cauchy. Overall, these kernels span a broad range of prior regularity, from relatively rough latent behavior to very smooth reconstructions, allowing us to assess the sensitivity of the results to the assumed smoothness class \cite{o2021elucidating}. Their explicit forms are summarized in Appendix~\ref{app:kernels}.

To connect the latent process to the observed Hubble measurements, we introduce an effective uncertainty \cite{2010arXiv1003.1315R}
\begin{equation}
\sigma_i^{\rm eff}
=
\sqrt{\sigma_{H,i}^2+\sigma_{\rm jit}^2},
\end{equation}
where $\sigma_{\rm jit}$ is an additional jitter term that absorbs possible underestimation of the reported errors. We assign
\begin{equation}
\begin{aligned}
\sigma_{\rm jit} &\sim \mathrm{HalfNormal}(\tau),\\
\tau &=
\operatorname{median}\!\left(\{\sigma_{H,i}\}_{i=1}^{n}\right).
\end{aligned}
\end{equation}

To reduce sensitivity to outliers and mild non-Gaussian residuals, we adopt a Student-$t$ likelihood \cite{student-t2009},
\begin{equation}
H_i\mid \vartheta,\mathcal M
\sim
\mathrm{Student}\text{-}t
\!\left(
\nu,\mu_i^{(\mathcal M)},\sigma_i^{\rm eff}
\right),
\end{equation}
where \(\mu_i^{(\mathcal M)}\) is the location parameter at redshift \(z_i\), \(\sigma_i^{\rm eff}\) is the scale, \(\nu\) is the degrees of freedom, and \(\vartheta\) denotes the full set of latent variables and hyperparameters. We parameterize the tail thickness as $\nu=2+\nu_+$, with $\nu_+ \sim \mathrm{Exponential}(\lambda_\nu)$ and $\lambda_\nu = \frac{1}{30}$.

Given the heterogeneous quality of current OHD, mild non-Gaussian residuals or occasional discordant points may remain after propagating the reported errors. The Student-$t$ form therefore provides a robust compromise: it down-weights tail events when needed, while approaching Gaussian behavior when the inferred $\nu$ is large.
\subsubsection{Two latent reconstruction strategies}
The two reconstruction methods differ mainly in the quantity to which the GP prior is assigned.

In Method~A, the latent process is assigned to the normalized dark-energy density itself, so that
\begin{equation}
f_i=g_i.
\end{equation}
In this direct latent-\(f\) branch, positivity of \(f(z)\) is not imposed by construction; the constrained log-\(f\) variant discussed in Section~\ref{sec:logf_branch} is used to assess the impact of enforcing \(f(z)>0\). We do not additionally impose \(f(0)=1\) as a hard GP anchor.

The Hubble rate entering the likelihood is then obtained from the Friedmann relation
after applying a smooth softplus--floor positivity mapping to the Friedmann argument
\cite{dugas2001incorporating},
\begin{equation}
\begin{split}
\mu_i^{(A)}
&=
H_0\Biggl\{
\mathcal{S}\!\Bigl[
\Omega_{m0}(1+z_i)^3+\Omega_{k0}(1+z_i)^2 \\
&\qquad
+\left(1-\Omega_{m0}-\Omega_{k0}\right)f_i
\Bigr]
\Biggr\}^{1/2},
\end{split}
\label{eq:mu_A}
\end{equation}
where \(\mathcal{S}\) is a smooth positivity-preserving mapping with a small
positive floor.

Thus, Method~A uses a direct latent representation of \(f(z)\). The softplus--floor mapping is used only as a numerical regularization of the Friedmann argument before taking the square root, ensuring real-valued \(H(z)\) predictions for all sampled latent configurations. It is not an additional positivity constraint on \(f(z)\); the impact of enforcing \(f(z)>0\) is assessed separately in the log-\(f\) branch. 

In Method~B, the latent process is assigned directly to the Hubble expansion rate,
\begin{equation}
\mu_i^{(B)}=g_i.
\end{equation}
In this case, the likelihood is built directly from the latent \(H(z)\) process, without imposing an \(H_0\) anchor; \((H_0,\Omega_{m0},\Omega_{k0})\) enter mainly when mapping the reconstructed \(H(z)\) to \(f(z)\), \(w(z)\), and \(Om(z)\) in post-processing.  Consequently, Method~B gives identical PSIS-LOO predictive quantities for all three external-prior sets, so Table~\ref{tab:model_comparison_psis} lists Method~B only once. This equivalence applies only to the observed-\(H(z)\) likelihood; the derived dark-energy quantities and \(Om(z)\) summaries remain prior dependent.

The observed-data likelihood is therefore evaluated in $H$ space in both methods; the difference lies in the latent representation. Using the same observed-$H$ likelihood in both branches keeps the predictive target fixed. This allows differences in PSIS-LOO and in the derived dark-energy constraints to be interpreted as consequences of latent representation and uncertainty propagation, rather than as artifacts of differing likelihood definitions.

\subsubsection{Prediction and derived quantities}
Given posterior samples of the latent variables and kernel hyperparameters, the GP defines predictive distributions for the reconstructed function at arbitrary redshifts $\mathbf z_*$. For the differentiable kernels considered here, the GP prior induces a joint Gaussian distribution for the function values and their derivatives at arbitrary redshifts \(\mathbf z_*\). The derivative covariance expressions are summarized in Appendix~\ref{app:gp_derivatives}.

For each posterior draw $(\mathbf g^{(s)},\eta^{(s)})$, the conditional distribution of the function and its derivative at $\mathbf z_*$ is
\begin{equation}
\begin{bmatrix}
\mathbf g_* \\
\mathbf g_*'
\end{bmatrix}
\Big|
\mathbf g^{(s)},\eta^{(s)},\mathcal M
\sim
\mathcal N
\left(
\begin{bmatrix}
\mathbf m_*^{(s)} \\
\mathbf m_*^{\prime(s)}
\end{bmatrix},
\mathbf\Sigma_*^{(s)}
\right),
\end{equation}
where $\mathbf g_*$ and $\mathbf g_*'$ denote the reconstructed function values and derivatives at $\mathbf z_*$, $\mathbf m_*^{(s)}$ and $\mathbf m_*^{\prime(s)}$ are the predictive means, and $\mathbf\Sigma_*^{(s)}$ is the predictive covariance matrix.

These posterior draws are then propagated directly to the derived dark-energy quantities. In Method~A, the GP yields $f(z)$ and $f'(z)$, from which $w(z)$ is obtained through Eq.~\eqref{eq:w_of_f}. In Method~B, the GP yields $H(z)$ and $H'(z)$, which are mapped to $f(z)$ and $f'(z)$ through Eqs.~\eqref{eq:f_of_H} and \eqref{eq:fprime_from_H}, and then to $w(z)$ through Eq.~\eqref{eq:w_of_f}. The reconstructed functions and derived dark-energy summaries are then obtained from the propagated posterior samples.

\subsection{Model comparison and pointwise PSIS-LOO diagnostics}
\label{sec:psisloo}

We assess predictive performance using the leave-one-out expected log pointwise predictive density, ELPD$_{\mathrm{LOO}}$ \cite{vehtari2016bayesian,vehtari2017practical,vehtari2024pareto,gelman2014understanding},
\begin{equation}
\mathrm{ELPD}_{\mathrm{LOO}}(\mathcal M)
=
\sum_{i=1}^{n}
\log p(H_i\mid H_{-i},\mathcal M),
\label{eq:elpd_loo}
\end{equation}
where $H_{-i}$ denotes the dataset with the $i$th observation removed. Because exact LOO would require refitting each model \(n\) times, we use Pareto-smoothed importance sampling leave-one-out cross-validation (PSIS-LOO), an efficient approximation to exact LOO whose reliability is monitored through Pareto-\(k\) diagnostics \cite{vehtari2017practical}. The corresponding estimator is
\begin{equation}
\widehat{\mathrm{ELPD}}_{\mathrm{PSIS\text{-}LOO}}(\mathcal M)
=
\sum_{i=1}^{n}
\log \widehat p_{\mathrm{LOO},i}^{\mathrm{PSIS}},
\label{eq:psis_loo}
\end{equation}
where $\widehat p_{\mathrm{LOO},i}^{\mathrm{PSIS}}$ is the PSIS approximation to the leave-one-out predictive density for the $i$th datum. Details of the importance-sampling construction are summarized in Appendix~\ref{app:psis}.

In addition to the global summary ELPD$_{\mathrm{LOO}}$, we explicitly examine the pointwise PSIS diagnostics. The reliability of the PSIS approximation is monitored using the Pareto tail-shape diagnostic $\hat k_i$ for each data point. As a rough guide, $\hat k_i<0.5$ corresponds to a stable importance-sampling regime, $0.5\le \hat k_i<0.7$ is usually still acceptable but indicates increasingly influential observations, and $\hat k_i\ge 0.7$ warns that the PSIS approximation may become unreliable \cite{vehtari2017practical}. In the present analysis, the pointwise $\hat k_i$ values are therefore used not only as a technical reliability check, but also as a diagnostic of whether apparent predictive differences are broadly distributed across the sample or instead driven by a small number of OHD points.

For both Method~A and Method~B, the pointwise log-likelihood is evaluated in the observed $H(z)$ space. This ensures a common predictive target across latent constructions and kernels. Rather than fixing a single covariance kernel a priori, we first perform PSIS-LOO stacking over the kernel set using the full real-data sample \((z\leq2)\) in observed \(H(z)\) space \cite{vehtari2017practical}. This initial stacking step defines the kernel combination used for the real-data reconstruction reported in Table~\ref{tab:model_comparison_psis}. In all subsequent analyses \((z\leq1)\), \((z\leq1.5)\), the A-log-\(f\) branch, and the mock tests, we keep this kernel-stacking prescription fixed and do not re-run kernel selection, comparison, or stacking. This choice keeps the latent-space setup and predictive reference consistent across analyses, so that differences can be attributed to the data range or analysis scenario rather than to kernel reselection.

To compare two models, we report
\begin{equation}
\Delta \mathrm{ELPD}_{\mathrm{LOO}}
=
\mathrm{ELPD}_{\mathrm{LOO}}(\mathcal M_1)
-
\mathrm{ELPD}_{\mathrm{LOO}}(\mathcal M_2),
\end{equation}
together with the associated standard error. We also inspect the pointwise difference
\begin{equation}
\Delta \mathrm{loo}_i
=
\mathrm{loo}_{i,1}-\mathrm{loo}_{i,2},
\end{equation}
where \(\mathrm{loo}_{i,m}\) denotes the pointwise LOO log predictive contribution of model \(\mathcal M_m\), so that
\begin{equation}
\begin{aligned}
\Delta \mathrm{ELPD}_{\mathrm{LOO}}
&=\sum_{i=1}^n \Delta \mathrm{loo}_i,\\
SE\!\left(\Delta \mathrm{ELPD}_{\mathrm{LOO}}\right)
&=\sqrt{\,n\,\operatorname{Var}_i\!\left(\Delta \mathrm{loo}_i\right)}.
\end{aligned}
\end{equation}
We further define the normalized contrast
\begin{equation}
Z_{\Delta}
=
\frac{\Delta \mathrm{ELPD}_{\mathrm{LOO}}}{SE\!\left(\Delta \mathrm{ELPD}_{\mathrm{LOO}}\right)}.
\end{equation}
There is no universal significance threshold for $\Delta \mathrm{ELPD}_{\mathrm{LOO}}$; its interpretation depends on the scale of the predictive difference together with its uncertainty \cite{vehtari2017practical,vehtari2024pareto,looFAQ2024}.
As a practical reporting rule \cite{looFAQ2024}, we treat $|\Delta \mathrm{ELPD}_{\mathrm{LOO}}|<4$ as a small predictive difference, while \(|\Delta \mathrm{ELPD}_{\mathrm{LOO}}|>4\) is treated as potentially relevant. In both cases, interpretation is based jointly on $\Delta \mathrm{ELPD}_{\mathrm{LOO}}$, $SE(\Delta)$ (or $Z_\Delta$), and pointwise $\Delta \mathrm{loo}_i$. This pointwise decomposition localizes where the predictive performance of the two reconstructions differs across the dataset.

In the comparisons below, Method~B is used as the reference model, since it corresponds to the conventional GP reconstruction applied directly to $H(z)$. 
Because our goal is to isolate the methodological impact of latent representation under a common predictive target, we report A-versus-B differences relative to this conventional $H$-space reference and interpret them as representation-driven differences.
\subsection{Mock-data construction and validation strategy}
\label{sec:mock_setup}

In addition to the real OHD analysis, we validate the full reconstruction pipeline on OHD-like mock data with known truth \cite{moresco2022unveiling,jimenez2002constraining,tomasetti2023new}. The mock study is designed to answer a more specific question raised by the real-data analysis: whether the residual A/B ambiguity is driven primarily by the reconstruction methodology or by the limited constraining power of the current CC subset of OHD, especially in the sparse high-redshift tail. We therefore use the mocks both to test signal recovery under controlled conditions and to determine how the A/B discrepancy changes when the high-redshift tail is made more informative.

We consider two truth models. The first is a fiducial \(\Lambda\)CDM truth model, for which $w(z)=-1$ and $f(z)=1$. The second is a mild Chevallier--Polarski--Linder (CPL) evolving model \cite{chevallier2001accelerating,linder2003exploring},
\begin{equation}
w_{\rm true}(z)=w_0+w_a\frac{z}{1+z},
\end{equation}
where we adopt \((w_0,w_a)=(-0.9,-0.4)\) in this work, with corresponding dark-energy density
\begin{equation}
f_{\rm true}(z)=(1+z)^{3(1+w_0+w_a)}
\exp\!\left[-3w_a\frac{z}{1+z}\right].
\end{equation}
The mock expansion history \(H_{\rm true}(z)\) is generated using the same
late-time Friedmann mapping as in the main analysis, with a fixed fiducial
background \((H_{0,\rm fid},\Omega_{m0,\rm fid},\Omega_{k0,\rm fid})
=(67.4,0.315,0.001)\). The resulting mock data sets are then analyzed under
the external prior sets listed in Table~\ref{tab:external_priors}.

For each truth model, mock data are generated directly in $H$ space under two OHD-like sampling configurations. In the observed-sampling configuration, the mock data follow the observed OHD redshift positions and reported $\sigma_H$ values. For the denser high-\(z\) configuration, the Table~\ref{tab:ccdata} sampling is kept unchanged at \(z<1.5\); the \(z\ge1.5\) tail is doubled in sampling density by linear interpolation up to \(z_{\max}=2\), and the corresponding \(\sigma_H\) values are multiplied by 0.5. We generate \(N_{\rm mock}=20\) independent noise realizations for each truth model and sampling configuration. For each realization, Gaussian random noise with standard deviation
\(\sigma_H\) is added pointwise according to the adopted uncertainty pattern.

Each mock realization is analyzed with the same inference procedure as the real data, including the same Method~A/Method~B definitions, external prior sets, and posterior propagation to $f(z)$, $w(z)$, and $Om(z)$. In this way, the mock study tests the recovery properties of the end-to-end reconstruction procedure rather than those of a simplified surrogate problem.

\subsection{Cross-method discrepancy diagnostics}
\label{sec:discrepancy_metrics}

Beyond predictive comparison in observed \(H\) space, we quantify how much the two latent constructions differ in reconstructed dark-energy space. We focus on \(f(z)\), since it is the primary reconstructed dark-energy quantity and is less singular than the derived \(w(z)\).
For each prior setting, we compare posterior draws of
\begin{equation}
\Delta f(z)=f_A(z)-f_B(z).
\end{equation}
To avoid assuming one-to-one correspondence between Markov-chain samples, we form Monte Carlo pairs by independently drawing posterior samples from Method~A and Method~B and then computing their differences.
At each redshift, we compute Cohen's \(d\),
\begin{equation}
d(z)=\frac{\mu_A(z)-\mu_B(z)}{\sqrt{\left[\sigma_A^2(z)+\sigma_B^2(z)\right]/2}},
\end{equation}
where $\mu_{A,B}$ and $\sigma_{A,B}$ are posterior means and standard deviations.
For descriptive magnitude interpretation, we use the conventional guide
\(|d|\approx0.2/0.5/0.8\) (small/medium/large) \cite{cohen1988statistical,cohen1992power},
while retaining the sign of \(d\) to indicate direction.

We summarize directional certainty by
\begin{equation}
pd(z)=\max\!\left\{P\!\left(\Delta f(z)>0\right),\,P\!\left(\Delta f(z)<0\right)\right\},
\end{equation}
estimated from paired posterior draws \cite{makowski2019indices}.
Values near $0.5$ indicate no directional preference, while values near $1$ indicate stable direction. As a rough descriptive reference, we regard $pd>0.95$ as indicating strong directional stability.
For redshift interval $I=[z_0,z_1]$, we define the interval-averaged absolute discrepancy
\begin{equation}
A_I=\frac{1}{|I|}\int_I |\Delta f(z)|\,\mathrm dz.
\end{equation}
Given a ROPE threshold $\delta$, we report
\begin{equation}
P(A_I<\delta),
\end{equation}
which quantifies the posterior support for the cross-method difference being practically small over the interval $I$ \cite{kruschke2018rejecting}.
We evaluate this quantity for
$I\in\{[0,1],[1,1.5],[1.5,2],[0,2]\}$.
The main-text ROPE threshold, \(\delta=3\), is a dimensionless threshold for \(A_I\). It is interpreted as a data-resolution or detectability scale motivated by the null fiducial-\(\Lambda\)CDM mock behavior, rather than as a physical-equivalence scale; we also report sensitivity checks for \(\delta=1\) and \(\delta=5\).

\section{Results}
\label{sec:results}

Before detailing each component, we emphasize that the results below should be interpreted primarily as a study of reconstruction robustness under current OHD quality, rather than as evidence that the real data favor a specific reconstruction method or require dynamical dark energy.

At the predictive level, the A/B difference is modest, sign-mixed, and prior dependent, with \(\Delta \mathrm{ELPD}_{\mathrm{LOO}}\) remaining below unity in magnitude. All reported reconstructions satisfy \(\max(\hat{k})<0.7\). For the reconstructed dark-energy quantities, Methods~A and~B remain broadly consistent, while latent-choice discrepancy becomes more visible toward the high-redshift tail; no method-independent departure from \(\Lambda\)CDM is robustly established by the current CC subset of OHD. 
 
For the log-\(f\) extension, this branch is technically feasible with the current CC subset of OHD, but because it enforces \(f(z)>0\) by construction, we treat it as a constrained sensitivity test rather than as our primary reconstruction. While it suppresses zero crossings, it can also alter the inferred \(w(z)\) evolution by removing posterior support with \(f(z)<0\). In the mock-data analysis, the main leverage for reducing A/B ambiguity comes from improved high-redshift sampling and precision: for mocks with injected mild evolution, the framework tracks this behavior, and the discrepancy between the two latent constructions decreases once the \(z\gtrsim1.5\) tail is made more informative. Overall, these results indicate that the dominant present limitation lies in sparse and prior-sensitive high-redshift OHD, rather than in a decisive real-data preference for either latent construction.

\subsection{Predictive model comparison}
\label{sec:results_model_comp}
We first examine the kernel dependence, with the PSIS-LOO comparison summarized in Table~\ref{tab:model_comparison_psis}. Across the kernel set on the full real-data \(z\leq2\) sample, ELPD differences are modest and the stacking weights are concentrated on a single dominant kernel in each reported case, so the stacked reconstruction is effectively identical to the corresponding top-ranked kernel. We then fix these kernels for all subsequent analyses, rather than using an explicit stacked average in later sections. In practice, the main-text figures use exactly the top-ranked kernel selected by the full-sample exploratory PSIS-LOO scan for each reported prior--method case. The full kernel-by-kernel comparison is given in Table~\ref{tab:psis_table} of Appendix~\ref{app:psis}. All downstream cases, including the \(z\)-truncated analyses, the A-log-\(f\) extension, and the mock-data tests, then reuse the same selected kernels.


The resulting PSIS-LOO differences are small and prior dependent:
\begin{equation}
\begin{split}
\Delta \mathrm{ELPD}_{\mathrm{LOO}}
&\equiv
\mathrm{ELPD}_{\mathrm{LOO}}(\mathrm{A})
-
\mathrm{ELPD}_{\mathrm{LOO}}(\mathrm{B})\\
&\in [-0.90,\,+0.35].
\end{split}
\end{equation}

\begin{table}[!htbp]
\centering
\small
\begingroup
\setlength{\tabcolsep}{1.4pt}
\caption{PSIS-LOO model comparison for the reported A/B reconstructions.
Method~B appears once because its PSIS-LOO quantities in observed \(H(z)\) space
are identical for the three prior sets. \(\Delta\)ELPD is computed relative to Method~B.}
\label{tab:model_comparison_psis}
\begin{tabular}{@{}llccc@{}}
\toprule
Case & Kernel & ELPD\(\pm\)SE & \(\Delta\)ELPD & \(\hat{k}_{\max}\) \\
\midrule
A: Planck2018   & Cauchy          & \(-153.13\pm3.68\) & \(+0.35\) & 0.42 \\
A: DESI--BBN & Mat\'ern \(5/2\) & \(-153.39\pm3.69\) & \(+0.09\) & 0.44 \\
A: SH0ES--Pantheon   & Cauchy          & \(-154.38\pm4.02\) & \(-0.90\) & 0.68 \\
B: Same for all priors   & Mat\'ern \(9/2\) & \(-153.48\pm3.96\) & ref.     & 0.59 \\
\bottomrule
\end{tabular}
\endgroup
\end{table}

Across priors, \(\Delta \mathrm{ELPD}_{\mathrm{LOO}}\) changes sign while remaining within this narrow range, so the PSIS-LOO comparison does not provide a decisive preference between the two latent constructions. We therefore treat Methods~A and~B as comparably supported by the current CC subset of OHD, and use the subsequent differences in \(f(z)\) and \(w(z)\) primarily to assess sensitivity to the choice of latent representation. Among the reported reconstructions, the SH0ES--Pantheon prior set yields the largest Pareto tail-shape estimate, with \(\max_i(\hat k_i)=0.68\) in Table~\ref{tab:model_comparison_psis}. Although this remains below the usual caution threshold, it is the closest case to the regime in which a small number of observations can become influential under PSIS.

\begin{figure}[!htbp]
\centering
\includegraphics[width=0.99\columnwidth]{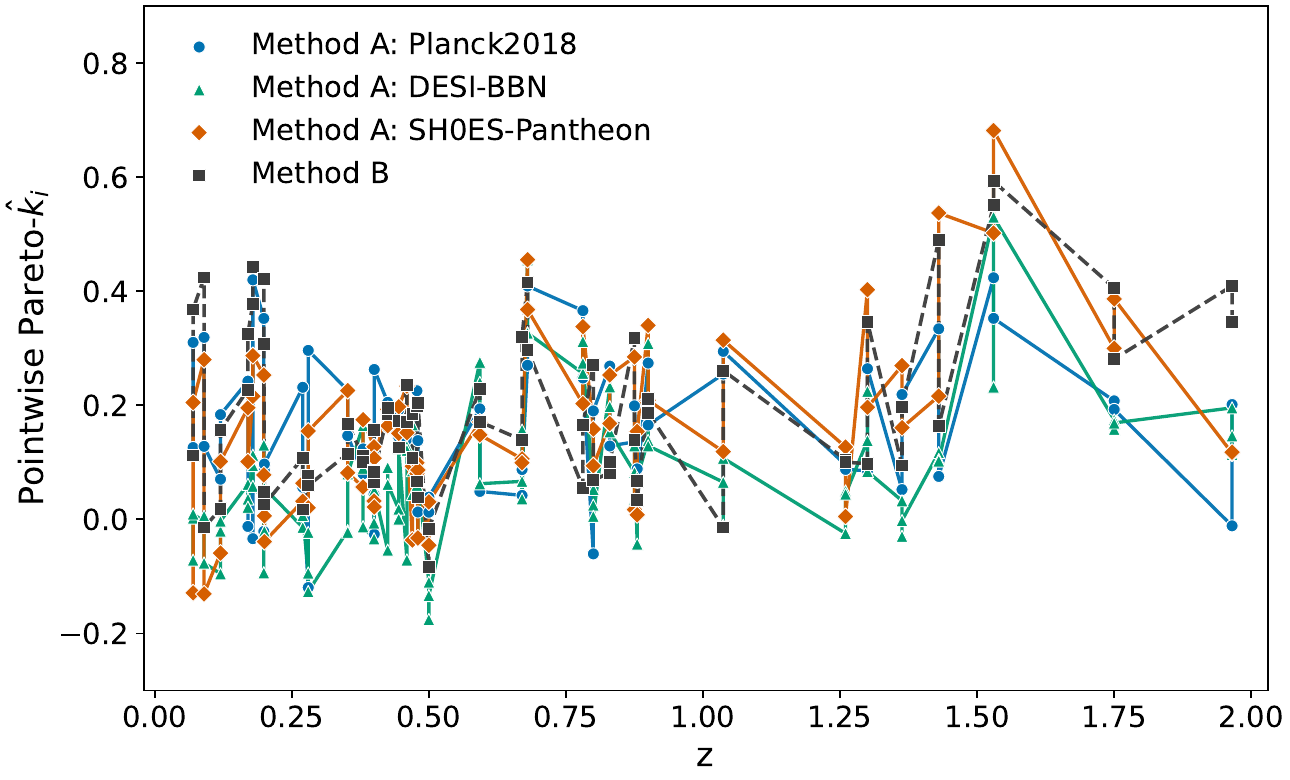}
\caption{
Pointwise Pareto-$\hat{k}_i$ diagnostics for the reconstructions.
The three colored curves show Method~A under the Planck2018, DESI--BBN, and SH0ES--Pantheon prior sets, while the black dashed curve shows Method~B, which is identical for all three prior sets in observed \(H(z)\) space.}
\label{fig:pointwise_pareto}
\end{figure}

Fig.~\ref{fig:pointwise_pareto} presents the pointwise Pareto-\(\hat{k}_i\) values for the reconstructions under comparison. For all reported reconstructions, \(\max_i(\hat{k}_i)<0.7\), indicating that the PSIS-LOO approximation remains acceptable in the main-text comparison. The pointwise behavior is nevertheless redshift dependent, with the largest \(\hat{k}_i\) values occurring toward the high-redshift tail. Across the broader kernel scan, including non-selected candidates, occasional localized \(\hat{k}_i>0.7\) values appear mainly at \(z\gtrsim1.5\), which motivates the dedicated mock tests with denser and more precise sampling in the high-redshift tail presented in Sec.~\ref{sec:results_mock}.

To localize the predictive differences, Fig.~\ref{fig:delta_loo_pointwise} in Appendix~\ref{app:additional_figures} shows the pointwise contribution
\begin{equation}
\Delta \mathrm{loo}_i \equiv \mathrm{loo}_{i,A}-\mathrm{loo}_{i,B}.
\end{equation}
The pointwise differences are small at low redshift and remain mixed at intermediate redshift. Positive \(\Delta \mathrm{loo}_i\) contributions become more visible in the high-redshift tail for the Planck2018 and DESI--BBN comparisons, whereas the SH0ES--Pantheon case remains mixed. Any net preference for Method~A is therefore localized and prior dependent, rather than a uniform improvement over the full sample.
Appendix~\ref{app:lml} reports the corresponding log marginal likelihood (LML) values. Across the reported reconstructions, the LML favors Method~A. However, because marginal-likelihood comparisons are generally more sensitive to prior specification, we treat the LML results as supplementary information rather than as the basis for our main comparative conclusions.

\subsection{Reconstruction of $f(z)$ and $w(z)$}
\label{sec:results_fw}
\begin{figure*}
\centering
\includegraphics[width=0.98\textwidth]{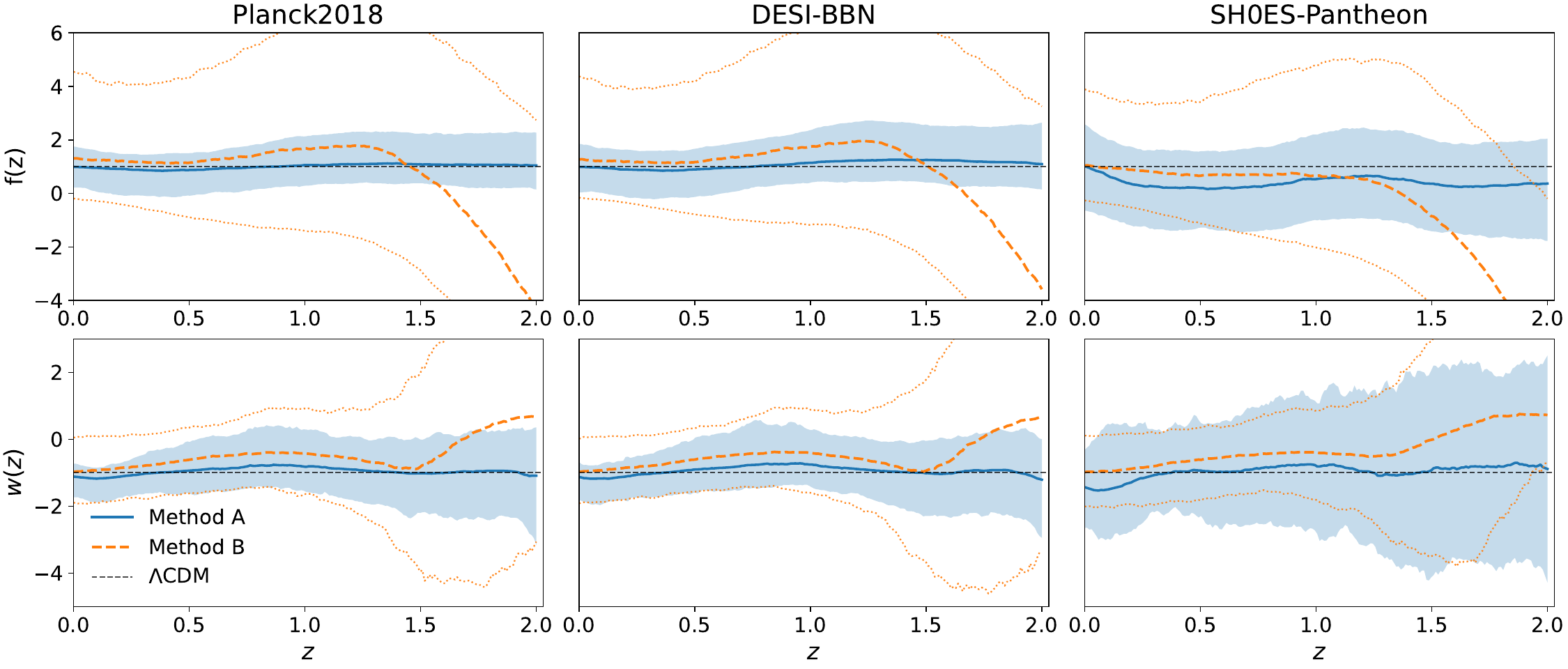}
\caption{
Comparison of \(f(z)\) and \(w(z)\) for Methods~A and~B using the full \(z\leq 2\) CC OHD sample.
Columns correspond to the Planck2018, DESI--BBN, and SH0ES--Pantheon prior sets; the top row shows \(f(z)\), and the bottom row shows \(w(z)\).
Method~A is shown by blue solid curves with blue shaded \(68\%\) credible bands.
Method~B is shown by orange dashed curves, with its \(68\%\) credible interval indicated by orange dashed bounds.
}
\label{fig:fw_main}
\end{figure*}

Fig.~\ref{fig:fw_main} collects the same-prior comparison between Methods~A and~B for the full $z\leq 2$ sample. The columns show the three external prior sets listed in Table~\ref{tab:external_priors}, while the two rows show $f(z)$ and $w(z)$. For all three prior sets, the reconstructed dark-energy quantities remain compatible with the $\Lambda$CDM expectations, $f(z)=1$ and $w(z)=-1$, within the current uncertainties over the full redshift range. The posterior bands from Methods~A and~B overlap substantially, so the current CC subset of OHD does not support a statistically meaningful or method-independent difference between the two latent constructions at the level of the reconstructed dark-energy quantities.

A clear methodological difference nevertheless remains in the credible-band width. Method~A generally yields tighter credible intervals for both $f(z)$ and $w(z)$ over most of the plotted range, whereas Method~B remains broader after the nonlinear mapping from $H(z)$ to the derived dark-energy quantities. This should be interpreted as a difference in posterior concentration rather than as evidence that one method is intrinsically more accurate.

The difference is most visible in the derived quantity $w(z)$ at higher redshift. Since $w(z)$ depends on both a derivative and a ratio, it is more sensitive to sparse sampling and finite measurement precision. Accordingly, the posterior uncertainty broadens substantially toward the high-redshift end in both methods, especially for $z\gtrsim1.5$, while remaining compatible with $w=-1$ within the inferred uncertainties. Prior dependence is also visible in Fig.~\ref{fig:fw_main}, most noticeably for the SH0ES--Pantheon case, but these shifts do not constitute robust evidence for departures from \(\Lambda\)CDM.

\subsection{Pole-crossing diagnostics and redshift-coverage checks}
\label{sec:pole_diag}

As discussed in Sec.~\ref{sec:theory}, $w(z)$ contains a formal pole when $f(z)=0$. This feature becomes relevant whenever the posterior support of $f(z)$ extends close to zero. To quantify its impact, we examine two posterior diagnostics: the fraction of draws satisfying \(f(z)<0\) and the fraction satisfying \(|w(z)|>10\), where the latter is used as a practical flag for extreme equation-of-state excursions rather than as a physical threshold.

\begin{figure}[!htbp]
\centering
\includegraphics[width=1\columnwidth]{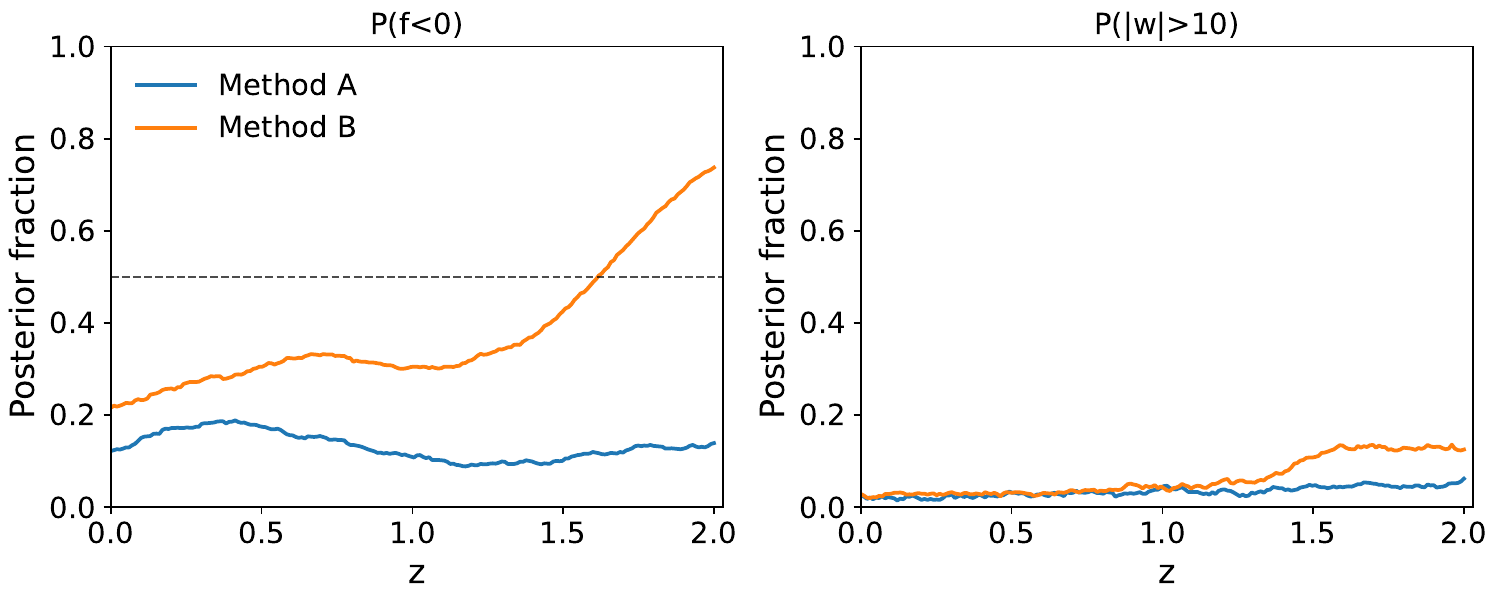}
\caption{
Posterior pole-crossing diagnostics for the Planck2018 prior set.
The plotted fractions quantify the posterior mass with $f(z)<0$ and with $|w(z)|>10$ as functions of redshift.
Both diagnostics remain small over most of the redshift range, showing that large excursions in $w(z)$ are driven by a minority of posterior draws.
}
\label{fig:pole_diagnostic}
\end{figure}

Fig.~\ref{fig:pole_diagnostic} shows these quantities for the Planck2018 prior set as a representative case. Both fractions remain small over most of the redshift range, indicating that \(f(z)=0\) crossings occur only in a minority of posterior realizations. The large excursions in $w(z)$ therefore arise from rare draws in which $f(z)$ approaches zero, rather than from the bulk posterior. This illustrates the role of the fully Bayesian sampling scheme: because inference is based on the full posterior rather than on a single reconstructed curve, such \(f(z)=0\) crossings are present but do not dominate the posterior medians or central credible bands. In this sense, the posterior summaries of $w(z)$ remain considerably more stable than a single plug-in reconstruction would suggest.

To assess sensitivity to redshift coverage, we repeat the same-prior comparison between Methods~A and~B on nested truncations with $z_{\max}=1$ and $z_{\max}=1.5$ (Appendix Fig.~\ref{fig:fw_trunc_appendix}). Relative to the full-sample result in Fig.~\ref{fig:fw_main}, posterior broadening in the derived $w(z)$ is most pronounced when the highest-redshift tail is retained and becomes milder after truncation.

Table~\ref{tab:discrepancy_interval_real} summarizes the interval-level ROPE diagnostics from pooled A/B real-data results under matched priors. Across intervals, the pooled median discrepancy increases from \(A_I=1.95\text{--}2.09\) in \(I=[0,1]\) to \(4.18\text{--}4.53\) in \(I=[1.5,2]\), while the posterior mass below the ROPE threshold decreases from \(P(A_I<3)=0.73\text{--}0.74\) to \(0.28\text{--}0.33\). The cross-prior ranges of \([q_{16},q_{84}]\) also widen toward high redshift, indicating less tightly constrained interval diagnostics in the sparsely sampled tail of the current CC subset of OHD. These pooled summaries are descriptive and do not imply a direct method-by-method ranking. In Sec.~\ref{sec:results_mock}, we present dedicated high-redshift mock experiments to further examine this regime.

\begin{table}[tp]
\centering
\begingroup
\caption{Interval-level ROPE diagnostics from pooled A/B real-data results under matched priors
(\(\Delta f=f_A-f_B\), \(\delta=3\)). Each entry reports the min--max over
Planck2018, DESI--BBN, and SH0ES--Pantheon; minima and maxima may come from
different priors across intervals and metrics. The column
``Range of [\(q_{16},q_{84}\)] across priors'' reports cross-prior endpoint ranges,
not a single 68\% credible interval from one posterior. Prior-specific values are listed in
Table~\ref{tab:discrepancy_appendix_full}.}

\label{tab:discrepancy_interval_real}
\begin{tabular}{@{}lccc@{}}
\toprule
Interval \(I\) & \(A_I\) median & Cross-prior range of \([q_{16}, q_{84}]\) & \(P(A_I<3)\) \\
\midrule
\([0,1]\)   & \(1.95\text{--}2.09\) & \([1.01\text{--}1.11,\ 3.84\text{--}4.01]\) & \(0.73\text{--}0.74\) \\
\([1,1.5]\) & \(2.80\text{--}2.91\) & \([1.20\text{--}1.30,\ 5.60\text{--}6.12]\) & \(0.52\text{--}0.54\) \\
\([1.5,2]\) & \(4.18\text{--}4.53\) & \([1.98\text{--}2.09,\ 7.48\text{--}8.08]\) & \(0.28\text{--}0.33\) \\
\([0,2]\)   & \(2.94\text{--}3.13\) & \([1.71\text{--}1.87,\ 4.90\text{--}5.07]\) & \(0.47\text{--}0.52\) \\
\bottomrule
\end{tabular}
\endgroup
\end{table}

\subsection{Feasibility of the log-$f$ reconstruction}
\label{sec:logf_branch}

In addition to the primary A/B branch, we test a log-\(f\) variant of Method~A, hereafter denoted A-log-\(f\), that places the GP prior on \(g(z)=\ln f(z)\) and maps back through \(f(z)=e^{g(z)}\). Using Eq.~\eqref{eq:w_of_f}, the dark-energy equation of state can then be rewritten as
\begin{equation}
\begin{split}
w(z)
&=-1+\frac{1+z}{3}\,\frac{f'(z)}{f(z)}=-1+\frac{1+z}{3}\,[\ln f(z)]'\\
&=-1+\frac{1+z}{3}\,g'(z).
\end{split}
\label{eq:w_of_logf}
\end{equation}
This makes explicit why the log-\(f\) parameterization is methodologically attractive: the derived \(w(z)\) is controlled directly by the derivative of the latent function, while the explicit \(1/f(z)\) factor in Eq.~\eqref{eq:w_of_f} is absorbed into the logarithmic reparameterization. The construction also enforces \(f(z)>0\) draw by draw, so configurations in which \(f(z)\) crosses zero, which drive the formal pole in \(w(z)\), are excluded by construction.

\begin{figure*}[tp]
\centering
\includegraphics[width=0.98\textwidth]{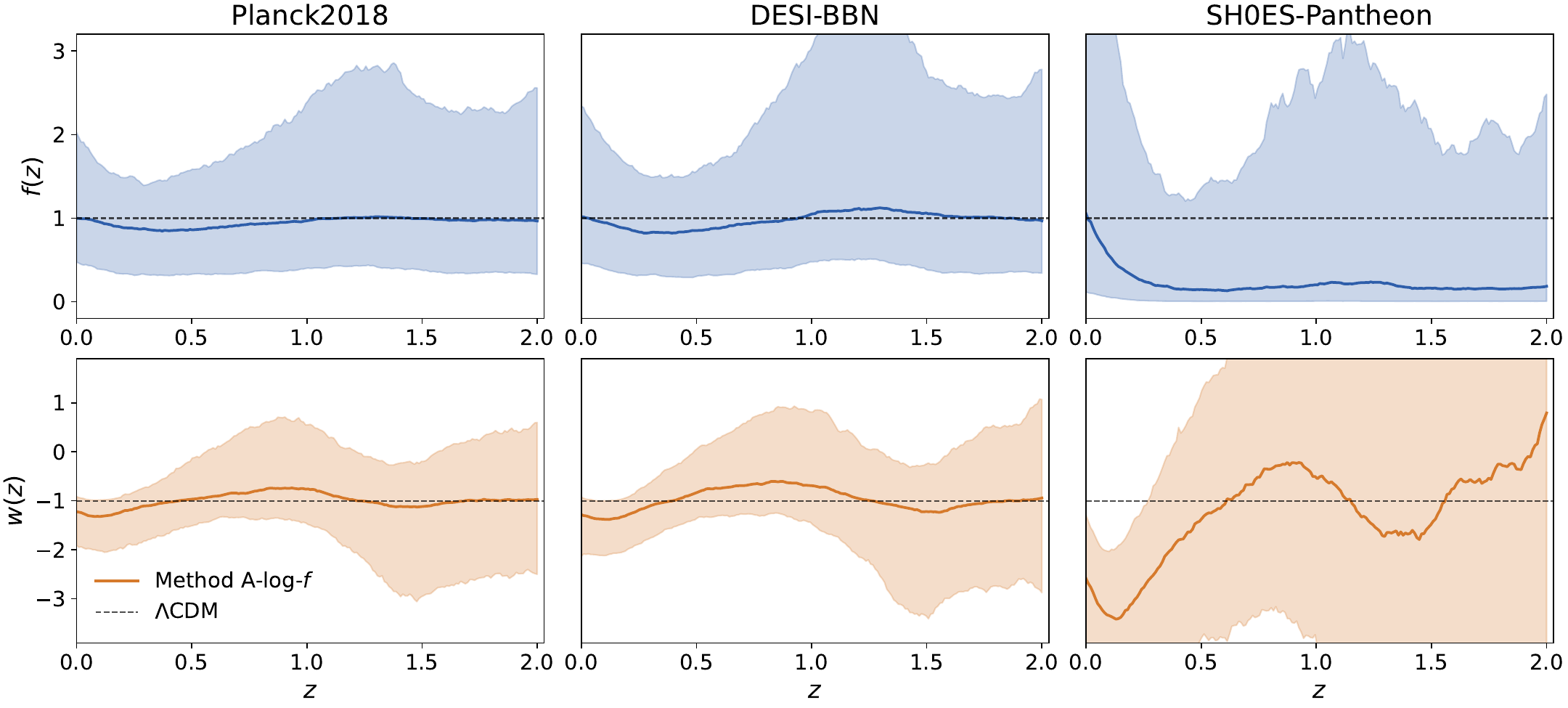}
\caption{
A-log-\(f\) reconstruction for the full $z\leq 2$ sample.
Columns show the Planck2018, DESI--BBN, and SH0ES--Pantheon prior sets; the top row shows $f(z)$ and the bottom row shows $w(z)$ for the log-\(f\) latent parameterization.
}
\label{fig:logf_diagnostics}
\end{figure*}

Fig.~\ref{fig:logf_diagnostics} shows that the log-\(f\) branch preserves the same broad cosmological conclusion as the A/B comparison: across all three prior sets, the reconstructed $f(z)$ and $w(z)$ remain compatible with $\Lambda$CDM within uncertainty. A prior-dependent boundary effect is nevertheless visible: for SH0ES--Pantheon, where the corresponding Method~A reconstruction is more sensitive to \(f(z)\) support near zero, enforcing \(f(z)>0\) in A-log-\(f\) truncates the lower tail around and below zero, potentially yielding a visibly different \(w(z)\) evolution pattern. For Planck2018 and DESI--BBN, posterior support near \(f(z)=0\) is much weaker, so the A-log-\(f\) and Method~A reconstructions remain closer.

The main methodological difference is that positivity of $f(z)$ is built in, so the most pathological configurations in which \(f(z)\) crosses zero are excluded by construction. Supplementary A-log-\(f\) reconstructions for the truncated $z_{\max}=1$ and $z_{\max}=1.5$ samples are shown in Appendix Fig.~\ref{fig:logf_facet_appendix}. The truncated-sample A-log-\(f\) results show the same qualitative behavior as Fig.~\ref{fig:logf_diagnostics}: the SH0ES--Pantheon prior set exhibits the largest prior-driven shift, while Planck2018 and DESI--BBN remain comparatively stable. This pattern indicates that, at the precision of the current CC subset of OHD, the inferred dark-energy reconstruction retains non-negligible dependence on the external \((H_0,\Omega_{m0},\Omega_{k0})\) prior sets. In the log-\(f\) branch, the posterior fraction with \(f(z)<0\) is identically zero by construction, while the large-\(|w|\) tail remains controlled in the reported posterior summaries. Quantitative interval-level summaries are reported in Table~\ref{tab:discrepancy_overview}. The full prior-expanded values and the corresponding \(Om(z)\) intervals are given in Appendix Tables~\ref{tab:discrepancy_appendix_full} and \ref{tab:om_all_models}, respectively. Together with \(\max(\hat{k})<0.7\) and the explicit suppression of configurations in which \(f(z)\) crosses zero, these results show that A-log-\(f\) is technically feasible with the current CC subset of OHD as a constrained sensitivity test. We do not regard A-log-\(f\) as a more model-independent reconstruction than Method~A, but rather as a constrained reparameterization whose built-in positivity prior changes the allowed posterior geometry.

\subsection{Mock-data validation and high-redshift sensitivity tests}
\label{sec:results_mock}
Using the mock-data construction described in Sec.~\ref{sec:mock_setup}, we examine four validation scenarios formed by combining two truth models with two OHD-like sampling configurations: the observed redshift configuration and a second configuration in which the \(z\ge 1.5\) tail is sampled more densely and with smaller \(\sigma_H\) values. These four cases serve two purposes: end-to-end recovery tests against known truth, and controlled sensitivity tests of how the derived dark-energy quantities change when the current high-redshift OHD tail becomes more informative.

\begin{figure*}[!tp]
\centering
\includegraphics[width=0.98\textwidth]{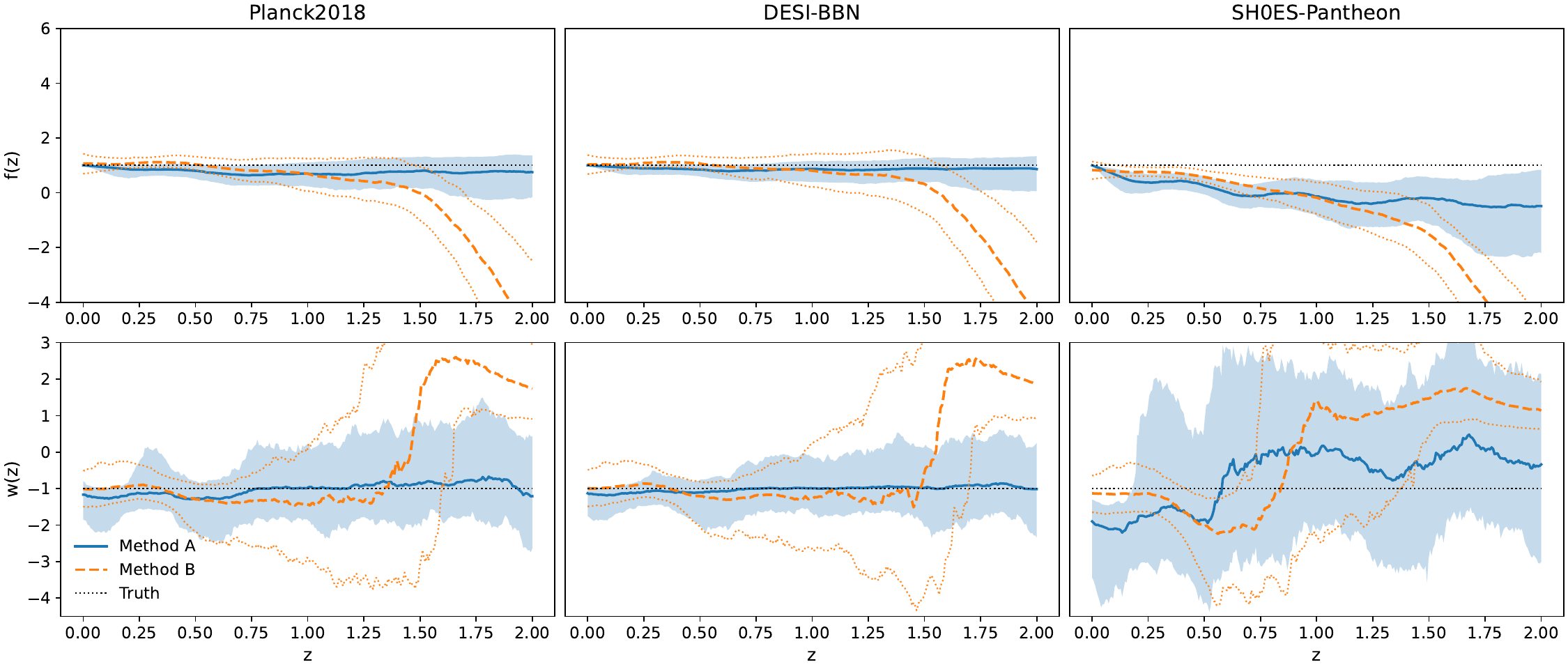}
\caption{
Representative mock-recovery example for the observed-sampling OHD-like configuration: fiducial \(\Lambda\)CDM truth.
Columns are Planck2018, DESI--BBN, and SH0ES--Pantheon; top row is \(f(z)\), bottom row is \(w(z)\). Method~A is shown by blue solid curves with blue shaded \(68\%\) credible bands, while Method~B is shown by orange dashed curves with orange dashed \(68\%\) bounds.
}
\label{fig:mock_baseline_lcdm}
\end{figure*}

The main purpose of these mock tests is not to establish a detection threshold in an absolute sense, but to determine whether the current real-data ambiguity is primarily method limited or data limited. We use the mock analysis as a controlled test of this question: if the Methods A/B discrepancy were mainly method limited, it should persist even when the truth is known and the high-redshift sampling is improved; if, instead, the discrepancy decreases and the injected signal becomes easier to recover once the high-redshift tail is strengthened, then the present ambiguity should be interpreted primarily as data limited. 

Fig.~\ref{fig:mock_baseline_lcdm} shows the fiducial-\(\Lambda\)CDM case for the observed-sampling OHD-like configuration, using the same three priors as in the real-data analysis. The complementary mild \(w_0w_a\) case and the two cases with denser and more precise high-redshift sampling are collected in Fig.~\ref{fig:mock_supp_examples} in Appendix~\ref{app:additional_figures}. The mock examples with improved high-redshift coverage show that the higher-precision tail reduces the posterior uncertainty of both \(f(z)\) and \(w(z)\), with the strongest narrowing occurring in the previously weakly constrained \(z\gtrsim1.5\) tail.

The figures serve as representative visual checks only. All quantitative conclusions from the mock-data analysis are drawn from ensemble-aggregated summaries. Discrepancy statistics are summarized in Table~\ref{tab:discrepancy_overview}, with prior-resolved values listed in Appendix Table~\ref{tab:discrepancy_appendix_full}. The corresponding \(Om(z)\) statistics are provided in Appendix Table~\ref{tab:mock_om_agg}.

Under the fiducial-\(\Lambda\)CDM truth, both methods can show apparent high-redshift deviations in the reconstruction of \(f(z)\), especially in the observed-sampling configuration. Under the SH0ES--Pantheon prior, the apparent deviations are more extended and can approach the \(1\sigma\) level over a broad redshift range. Compared with Method~B, Method~A tends to show weaker apparent evolution in \(f(z)\).

This contrast is more pronounced in the \(w(z)\) reconstruction. Method~A shows only a mild and localized \(1\sigma\) deviation at low redshift, whereas Method~B exhibits more extended apparent \(1\sigma\)-level deviations, particularly toward high redshift, consistent with its behavior in \(f(z)\).

For the mild \(w_0w_a\) truth, the behavior differs notably. Both methods show small, localized \(1\sigma\) deviations in \(f(z)\) at low redshift, while under the SH0ES--Pantheon prior the apparent deviations become more extended across redshift. In addition, Method~A can exhibit a visible low-redshift phantom-crossing pattern, which is absent in the fiducial-\(\Lambda\)CDM case. The corresponding results in the configurations with denser and more precise sampling in the high-redshift tail follow similar qualitative trends, while remaining closer overall to the injected truth.

We emphasize that these phantom-crossing features occur at the \(\sim1\text{--}2\sigma\) level and are not statistically decisive. Their dependence on the reconstruction method and prior choice suggests that such behavior may reflect reconstruction-method and prior dependence, highlighting the role of latent representation as a non-negligible source of uncertainty.

This behavior is qualitatively consistent with the greater sensitivity observed for the SH0ES--Pantheon prior set in the predictive diagnostics, where the relatively larger Pareto tail-shape estimate suggests that a small number of observations may become more influential. Such sensitivity can propagate into the reconstruction and amplify apparent deviations in the dark-energy quantities. A more quantitative characterization of these effects is given in Secs.~\ref{sec:results_discrepancy} and \ref{sec:consistency_diagnostics}.

\subsection{Cross-method discrepancy in reconstructed dark-energy space}
\label{sec:results_discrepancy}

We quantify the Methods A-versus-B difference directly in reconstructed $f(z)$ space using the diagnostics defined in Sec.~\ref{sec:discrepancy_metrics}.
To cover the full set of reported configurations, we perform a same-prior comparison between Methods~A and~B for all three prior settings in Table~\ref{tab:external_priors}.

\begin{figure}[tp]
\centering
\includegraphics[width=0.99\columnwidth]{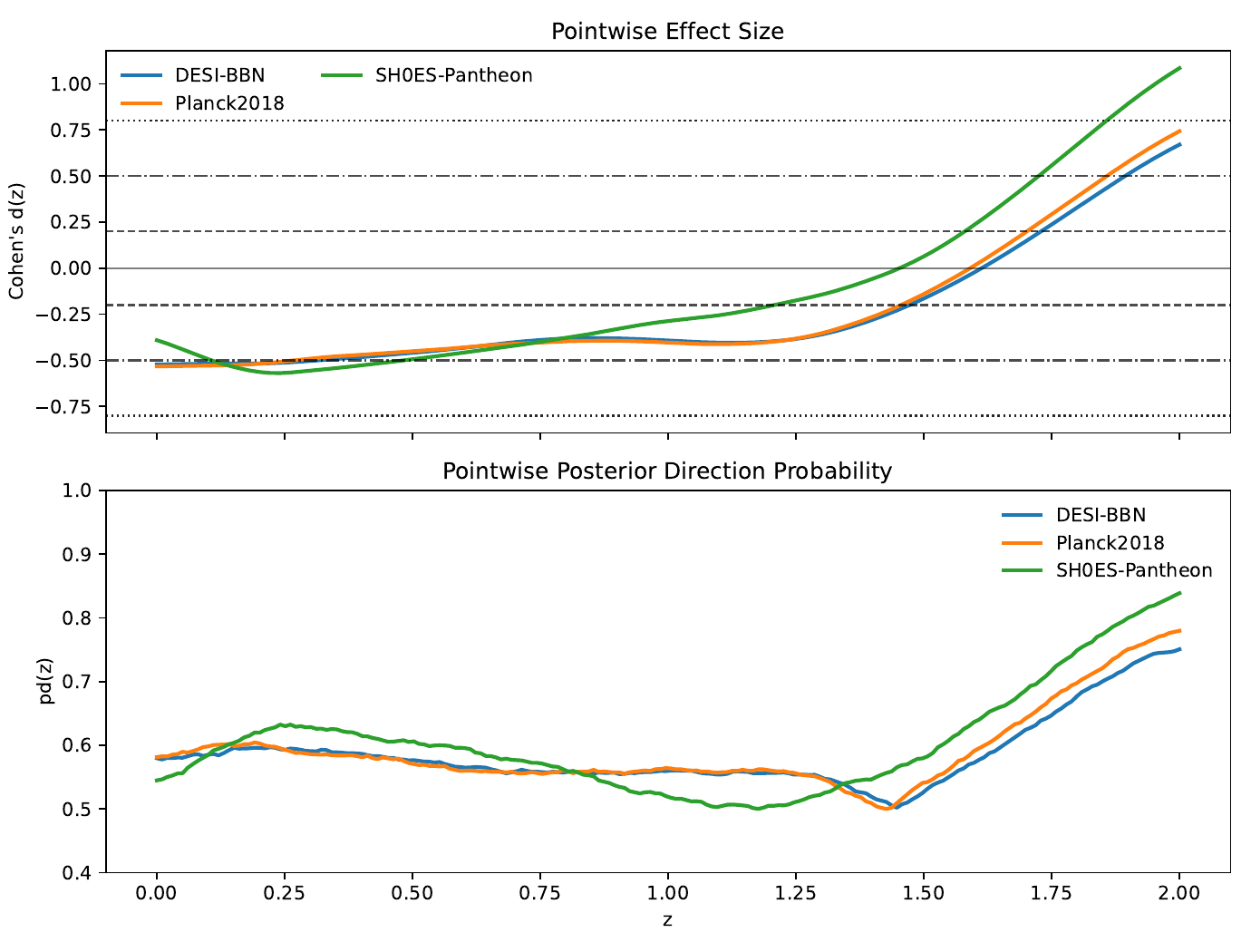}
\caption{
Pointwise cross-method discrepancy diagnostics in reconstructed $f(z)$ using same-prior comparison between Methods~A and~B across all three prior settings.
Top: Cohen's $d(z)$. Bottom: posterior direction probability $pd(z)$.
All curves use the sign convention $\Delta f=f_A-f_B$.
}
\label{fig:discrepancy_main}
\end{figure}
Fig.~\ref{fig:discrepancy_main} summarizes the pointwise results. The signed pointwise effect size is not sign-definite over the full grid: \(d(z)\in[-0.57,1.09]\) across the tested priors.
In magnitude, the pointwise effect sizes remain mostly moderate: across the prior-interval combinations summarized in Appendix Table~\ref{tab:discrepancy_appendix_full}, \(\max |d|\) ranges from \(0.29\) to \(1.09\).
The direction metric also remains moderate, with the corresponding \(\max pd\) values staying below \(0.84\), so none of the reported interval-level maxima enters the strong-direction regime \(pd>0.95\).
Thus, directional evidence is limited even when local effect sizes are nonzero. This indicates that the current data provide only limited discriminating power between the two methods at the pointwise level.

These interval summaries are consistent with the pointwise picture: \(A_I\) increases toward higher redshift, while \(P(A_I<3)\) decreases correspondingly, so posterior support for a practically small cross-method discrepancy is weakest in the \(z\in[1.5,2]\) tail, where the reconstructed \(w(z)\) is already most unstable. This indicates that cross-method discrepancies become more pronounced toward high redshift, but remain coupled to increased reconstruction uncertainty in this regime.

ROPE conclusions are threshold dependent by construction, so we also check $\delta=1$ and $\delta=5$. Across all priors and intervals, $P(A_I<\delta)$ spans $0.017$--$0.158$ for $\delta=1$ and $0.555$--$0.914$ for $\delta=5$. This sensitivity check shows that the redshift-dependent ordering of the intervals is stable across ROPE thresholds, whereas the absolute values of \(P(A_I<\delta)\) naturally depend on the chosen practical-equivalence scale.
Accordingly, \(\delta=3\) is read as a detectability-level criterion under current data quality, while \(\delta=1\) and \(\delta=5\) provide stricter/looser practical-equivalence sensitivity checks.

\begin{table*}[tp]
\centering
\small
\caption{Unified high-redshift discrepancy summary at \(I=[1.5,2]\) (\(\delta=3\)). For A/B rows (real and mock), \(\Delta f=f_A-f_B\); for A-log-\(f\) vs A, \(\Delta f=f_{A\text{-}\log f}-f_A\). Ranges are taken across the three prior settings; mock rows are aggregated over realizations within each prior. Here ``denser high-\(z\)'' denotes the mock setup in which the baseline configuration is retained but the \(z\ge1.5\) tail is sampled more densely and assigned smaller \(\sigma_H\) values.}
\label{tab:discrepancy_overview}
\begingroup
\setlength{\tabcolsep}{4pt}
\resizebox{\textwidth}{!}{%
\begin{tabular}{lcccc}
\toprule
Case & \(A_I\) median (range) & \(P(A_I<3)\) (range) & \(\max|d|\) (range) & \(\max pd\) (range) \\
\midrule
A vs B & \(4.18\text{--}4.53\) & \(0.28\text{--}0.33\) & \(0.67\text{--}1.09\) & \(0.75\text{--}0.84\) \\
A-log-\(f\) vs A & \(1.02\text{--}1.68\) & \(0.73\text{--}0.87\) & \(0.12\text{--}0.18\) & \(0.52\text{--}0.58\) \\
Mock (baseline, fiducial \(\Lambda\)CDM) & \(2.04\text{--}2.95\) & \(0.54\text{--}0.73\) & \(1.77\text{--}2.47\) & \(0.89\text{--}0.96\) \\
Mock (baseline, mild \(w_0w_a\)) & \(2.01\text{--}2.93\) & \(0.52\text{--}0.70\) & \(1.78\text{--}2.54\) & \(0.89\text{--}0.95\) \\
Mock (denser high-\(z\), fiducial \(\Lambda\)CDM) & \(0.97\text{--}1.44\) & \(0.90\text{--}0.98\) & \(1.08\text{--}1.95\) & \(0.77\text{--}0.91\) \\
Mock (denser high-\(z\), mild \(w_0w_a\)) & \(0.93\text{--}1.31\) & \(0.91\text{--}0.98\) & \(0.92\text{--}1.70\) & \(0.75\text{--}0.90\) \\
\bottomrule
\end{tabular}
}
\endgroup
\end{table*}

Table~\ref{tab:discrepancy_overview} provides a common quantitative reference for the main-text results: the real-data Method A/B discrepancy is concentrated in the high-\(z\) tail, the A-log-\(f\) branch remains closer to Method~A under the same priors, and the mock discrepancies are reduced in the sampling configurations with denser and more precise high-redshift tails. For each row, the ranges reported in all metric columns denote the minimum--maximum across the three prior settings; for mock rows, values are first aggregated over mock data sets within each prior and then summarized across priors.

For the mock A/B rows, the configuration with denser and more precise sampling in the high-redshift tail shifts \(A_I\) from \(2.01\text{--}2.95\) to \(0.93\text{--}1.44\), while \(P(A_I<3)\) increases from \(0.52\text{--}0.73\) to \(0.90\text{--}0.98\). This indicates, in these mock tests, that the present ambiguity in the reconstructed high-redshift dark-energy behavior is largely driven by limited OHD constraining power at \(z\gtrsim1.5\). Improved high-redshift coverage is therefore required before apparent evolution can be more reliably assessed relative to \(\Lambda\)CDM. The full prior-expanded interval tables for all real and mock cases are reported in Appendix Table~\ref{tab:discrepancy_appendix_full}.

\subsection{External-prior dependence and the \(Om(z)\) diagnostic}
\label{sec:consistency_diagnostics}
We next examine how the reported reconstructions depend on the external prior sets and quantify their consistency with $\Lambda$CDM using the $Om(z)$ diagnostic.
\subsubsection{Prior dependence of \(f(z)\) and \(w(z)\)}
The main-text figure shows the three prior sets side by side for the full \(z\leq 2\) analysis (Fig.~\ref{fig:fw_main}), while the nested truncations \(z\leq 1\) and \(z\leq 1.5\) are reported in Appendix Fig.~\ref{fig:fw_trunc_appendix}. The qualitative conclusion is unchanged across these cases: the current CC OHD do not require a departure from $f(z)=1$ or $w(z)=-1$.

The central posterior trends do, however, show non-negligible dependence on the external prior sets, especially in Method~A. This is most evident for the SH0ES--Pantheon case, where the posterior medians shift more visibly away from the Planck2018-based reconstruction. Method~B is less sensitive to this choice, although the derived $f(z)$ and $w(z)$ still vary because the mapping from $H(z)$ to dark-energy quantities depends on $(H_0,\Omega_{m0},\Omega_{k0})$. The difference between Method~A and A-log-\(f\) is likewise most visible for SH0ES--Pantheon, consistent with the stronger near-zero posterior support of \(f(z)\) under this set.

\subsubsection{The \(Om(z)\) diagnostic}

\begin{figure*}[tp]
\centering
\includegraphics[width=\textwidth]{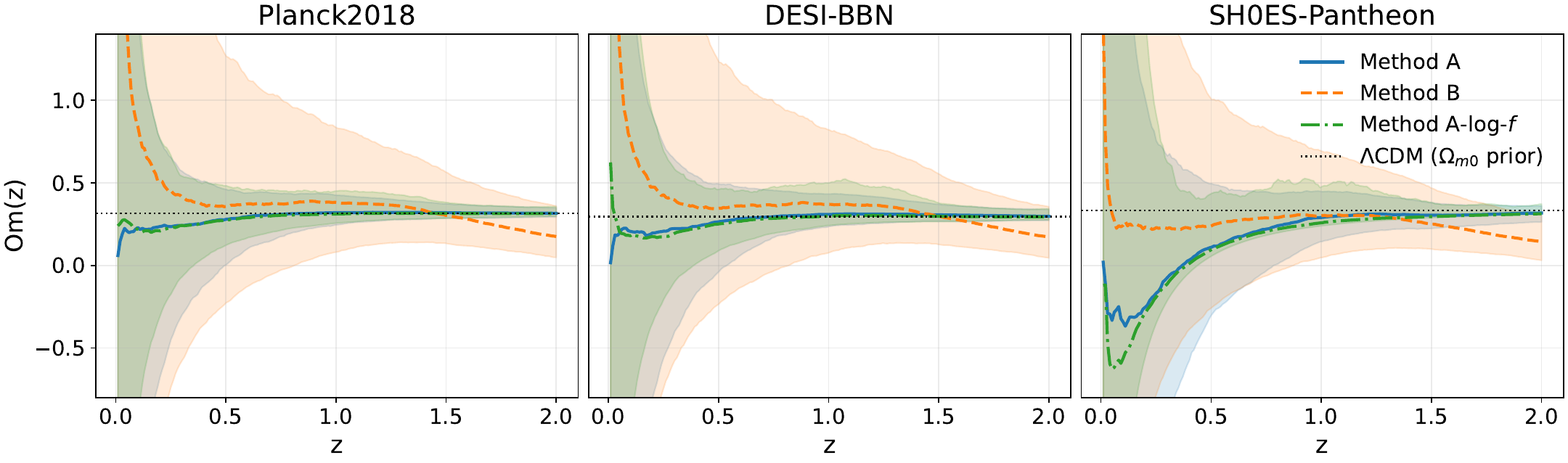}
\caption{
Posterior $Om(z)$ evolution comparison for Methods~A, B, and A-log-\(f\) in the full \(z\leq 2\) analysis.
Each panel corresponds to one prior set and shows median trends with \(68\%\) intervals for the three reconstruction branches.
}
\label{fig:omz_evolution}
\end{figure*}

Fig.~\ref{fig:omz_evolution} provides a direct visual summary of the prior-dependent \(Om(z)\) behavior. Consistent with the real-data entries in Appendix Table~\ref{tab:om_all_models}, the inferred \(\Delta Om_{0.5,1.0}\) and \(\Delta Om_{0.5,1.5}\) intervals remain compatible with zero at \(68\%\) for all methods and priors, while SH0ES--Pantheon exhibits the largest prior-driven shift. This indicates that the apparent redshift evolution in \(Om(z)\) is presently prior sensitive rather than a robust signature of departures from \(\Lambda\)CDM.

To further characterize the reconstructions, we evaluate \(Om(z)\) at representative redshifts and compute the two-point statistic \(\Delta Om_{ij}\) defined in Eq.~\eqref{eq:delta_om}. Posterior medians and central \(68\%\) credible intervals are listed in Appendix Table~\ref{tab:om_all_models}. Within the direct A/B comparison, Method~B remains systematically broader.

For mock data, we report \(Om(z)\) summaries aggregated over mock data sets rather than results from a single mock data set. Appendix Table~\ref{tab:mock_om_agg} summarizes \(\Delta Om_{0.5,1.0}\) and \(\Delta Om_{0.5,1.5}\) across mock data sets for each truth/configuration/prior/method combination, using the median and 16th--84th percentile of realization-level posterior medians. This aligns the \(Om(z)\) presentation with the discrepancy diagnostics aggregated over mock data sets.

Under this ensemble summary, denser and more precise sampling in the high-redshift tail reduces scatter across mock data sets and makes the injected mild-evolution response in \(\Delta Om\) more consistently visible in the ensemble medians, while the real-data \(Om(z)\) trend remains provisional and prior sensitive. The coverage fraction should be interpreted as a diagnostic rather than as a detection metric; even when the realization-aggregated median and 16th--84th percentile spread are compatible with the injected truth, a low \(C_{68}^{0.5,1.5}\) indicates possible undercoverage of the truth by the posterior \(68\%\) intervals for some prior--method combinations.
In summary, these results indicate that any apparent redshift evolution of \(Om(z)\) in the current real data is not yet statistically compelling and remains sensitive to both method and external-prior choice. The mock analysis confirms that the same statistic can respond to injected evolution when the signal and high-redshift information are sufficiently strong. The present limitation is therefore not the \(Om(z)\) diagnostic itself, but the combination of sparse high-redshift coverage and remaining dependence on external priors in the current CC subset of OHD.

\section{Conclusion} \label{sec:conclusion}

We studied dark-energy reconstruction from the current CC subset of OHD within a Bayesian Gaussian-process framework, comparing two latent constructions: Method~A, in which the GP prior is placed directly on \(f(z)\), and Method~B, in which the GP prior is placed on \(H(z)\) and the dark-energy quantities are derived in post-processing.

The PSIS-LOO comparison does not identify a decisive preference between the two latent constructions. Relative to the conventional latent-\(H\) Method~B reference, Method~A gives sign-mixed \(\Delta \mathrm{ELPD}_{\mathrm{LOO}}\) values in the range \([-0.90,\,0.35]\), indicating that any predictive preference is modest and prior dependent. The two latent constructions are therefore comparably supported by the current CC subset of OHD.
Method~A often yields tighter posterior bands in the derived dark-energy quantities, but this greater posterior concentration is not accompanied by a decisive predictive advantage.

At the same time, the reconstructed dark-energy quantities are not identical in the two formulations. The cross-method discrepancy is weakest at low redshift and becomes more visible toward the high-redshift tail, where the data are sparsest and the reconstructed \(w(z)\) is least stable. This indicates that latent-choice sensitivity in the current analysis is concentrated mainly in the poorly constrained \(z\gtrsim1.5\) regime rather than across the full redshift range.

From a cosmological standpoint, we interpret the reconstructed \(f(z)\), \(w(z)\), and \(Om(z)\) as remaining broadly compatible with \(\Lambda\)CDM for the current CC subset of OHD. We therefore regard any apparent redshift evolution in \(Om(z)\) as provisional and sensitive to external priors, rather than as method-independent evidence for departures from \(\Lambda\)CDM. The strongest shifts occur for the SH0ES--Pantheon prior set, showing that residual sensitivity to \((H_0,\Omega_{m0},\Omega_{k0})\) remains an important part of the current uncertainty budget.

We also explored an A-log-\(f\) extension, with a GP prior on \(\ln f(z)\). Because it enforces \(f(z)>0\) by construction, we interpret it as a constrained sensitivity test rather than as the main reconstruction used for inference. It suppresses configurations in which \(f(z)\) crosses zero, but can also modify the inferred \(w(z)\) evolution by removing posterior support with \(f(z)<0\). Importantly, it does not change the overall cosmological conclusion.

The OHD-like mock analysis helps separate effects caused by limited data quality from those caused by the reconstruction method itself. In the mild \(w_0w_a\) mocks, the \(Om(z)\) diagnostic responds to the injected evolution. Denser and more precise sampling in the high-redshift tail reduces the discrepancy between the two latent constructions and makes this response more consistently identifiable. The same denser and more precise high-redshift sampling also narrows the reconstructed uncertainty bands of \(f(z)\) and \(w(z)\), especially at \(z\gtrsim1.5\). At present, the main limitation comes from sparse high-redshift OHD coverage and residual sensitivity to external priors in the currently available data.

Overall, the present study indicates that latent-space choice is a relevant but secondary source of uncertainty in current OHD-based dark-energy reconstruction. At the present data quality, the dominant limitation comes instead from the sparse high-redshift tail and the remaining dependence on external priors. The latent-\(f\), log-\(f\), and cross-method comparison framework developed here is most useful for distinguishing methodological ambiguity from the limitations of the current data.

\begin{acknowledgments}
We thank Kang Jiao, Jing Niu, and Jian-kang Li for valuable discussions and helpful suggestions. This work was supported by the National SKA Program of China (No. 2022SKA0110202), the China Manned Space Program (CMS-CSST-2025-A01), and the China Scholarship Council (File No. 202506040057).
\end{acknowledgments}

\section*{Data Availability}
The data used in this study are compiled from published literature sources and are listed in Table~\ref{tab:ccdata}. No new observational data were generated. Mock data are generated from the procedures described in Sec.~\ref{sec:mock_setup}.

\appendix
\section{Kernel functions}
\label{app:kernels}

We use three covariance-kernel families in the GP reconstruction. For two
redshifts $z$ and $z'$, let $r=|z-z'|$, with correlation length
$\ell$ and amplitude $\sigma_g$.
The radial basis function kernel is
\begin{equation}
k_{\rm RBF}(z,z')
=
\sigma_g^2
\exp\!\left[-\frac{(z-z')^2}{2\ell^2}\right].
\end{equation}
The Mat\'ern family is written as
\begin{equation}
k_{\nu}(z,z')
=
\sigma_g^2
\frac{2^{1-\nu}}{\Gamma(\nu)}
\left(\frac{\sqrt{2\nu}\,r}{\ell}\right)^\nu
K_\nu\!\left(\frac{\sqrt{2\nu}\,r}{\ell}\right),
\end{equation}
where $K_\nu$ is the modified Bessel function of the second kind. In the analysis, we consider the half-integer cases
$\nu=3/2,\,5/2,\,7/2,\,9/2$, for which the kernel reduces to a
polynomial times an exponential.
The Cauchy kernel is
\begin{equation}
k_{\rm Cauchy}(z,z')
=
\sigma_g^2
\left[1+\frac{(z-z')^2}{\ell^2}\right]^{-1}.
\end{equation}

\section{Gaussian-process derivatives}
\label{app:gp_derivatives}

For the differentiable kernels considered in this work, if \(g(z)\sim \mathrm{GP}(0,k(z,z'))\), then the function and its first derivative are jointly Gaussian. Let \(\mathbf z\) denote the observed
redshifts and \(\mathbf z_*\) the test redshifts, and define
\begin{equation}
\mathbf K = k(\mathbf z,\mathbf z), ~ \mathbf K_* = k(\mathbf z,\mathbf z_*), ~ \mathbf K_*' = \partial_{z_*} k(\mathbf z,\mathbf z_*),
\end{equation}
together with
\begin{equation}
\begin{aligned}
\mathbf K_{**}^{00}
&= k(\mathbf z_*,\mathbf z_*),
&
\mathbf K_{**}^{01}
&= \partial_{z_*'} k(\mathbf z_*,\mathbf z_*'), \\
\mathbf K_{**}^{10}
&= \left(\mathbf K_{**}^{01}\right)^\top,
&
\mathbf K_{**}^{11}
&= \partial_{z_*}\partial_{z_*'} k(\mathbf z_*,\mathbf z_*').
\end{aligned}
\end{equation}
Conditioned on the latent values $\mathbf g$ at the observed
redshifts, the predictive distribution for the function and derivative
at $\mathbf z_*$ is
\begin{equation}
\begin{bmatrix}
\mathbf g_* \\
\mathbf g_*'
\end{bmatrix}
\Bigg|\mathbf g
\sim
\mathcal N
\left(
\begin{bmatrix}
\mathbf K_*^\top \mathbf K^{-1}\mathbf g \\
(\mathbf K_*')^\top \mathbf K^{-1}\mathbf g
\end{bmatrix},
\mathbf\Sigma_{*,\mathrm{joint}}
\right),
\end{equation}
with
\begin{equation}
\mathbf A_*=
\begin{bmatrix}
\mathbf K_*^\top \\
(\mathbf K_*')^\top
\end{bmatrix},
\qquad
\mathbf B_*=
\begin{bmatrix}
\mathbf K_* & \mathbf K_*'
\end{bmatrix},
\end{equation}
and
\begin{equation}
\mathbf\Sigma_{*,\mathrm{joint}}
=
\begin{bmatrix}
\mathbf K_{**}^{00} & \mathbf K_{**}^{01} \\
\mathbf K_{**}^{10} & \mathbf K_{**}^{11}
\end{bmatrix}
-\mathbf A_*\,\mathbf K^{-1}\mathbf B_*.
\end{equation}
These relations are used to obtain $f'(z)$ in Method~A and $H'(z)$ in
Method~B, which are then propagated to the derived equation-of-state
parameter $w(z)$.

\section{Importance-sampling formulation of PSIS-LOO}
\label{app:psis}

For model $\mathcal M$, the exact leave-one-out predictive density of
datum $H_i$ is

\begin{equation}
p(H_i \mid H_{-i}, \mathcal M)
=
\int
p(H_i \mid \vartheta,\mathcal M)
p(\vartheta \mid H_{-i},\mathcal M)\,
\mathrm d\vartheta,
\end{equation}

where $\vartheta$ denotes the full set of latent variables and
hyperparameters. Direct evaluation requires refitting the model $n$
times. Using posterior draws
$\{\vartheta^{(s)}\}_{s=1}^{S}\sim p(\vartheta\mid H,\mathcal M)$ from
the full-data posterior, the raw importance-sampling approximation is
\begin{equation}
\widehat p_{\mathrm{LOO},i}^{\mathrm{IS}}
=
\left[
\frac{1}{S}\sum_{s=1}^{S}
\frac{1}{p(H_i\mid \vartheta^{(s)},\mathcal M)}
\right]^{-1}.
\end{equation}

\begin{equation}
r_i^{(s)}
=
\frac{1}{p(H_i\mid \vartheta^{(s)},\mathcal M)}.
\end{equation}

Pareto-smoothed importance sampling (PSIS)
\cite{vehtari2017practical,vehtari2024pareto} stabilizes the upper
tail of the raw ratios $r_i^{(s)}$. With smoothed ratios
$\tilde r_i^{(s)}$ and normalized weights
\begin{equation}
\bar w_i^{(s)}=
\frac{\tilde r_i^{(s)}}{\sum_{s'=1}^{S}\tilde r_i^{(s')}},
\end{equation}
the PSIS estimate is
\begin{equation}
\widehat p_{\mathrm{LOO},i}^{\mathrm{PSIS}}
=
\sum_{s=1}^{S}\bar w_i^{(s)}\,
p(H_i\mid \vartheta^{(s)},\mathcal M).
\end{equation}
\begin{equation}
\widehat{\mathrm{ELPD}}_{\mathrm{PSIS\text{-}LOO}}
=
\sum_{i=1}^{n}\log \widehat p_{\mathrm{LOO},i}^{\mathrm{PSIS}}.
\end{equation}
The reliability of the approximation is monitored by the Pareto
tail-shape parameter $\hat k_i$; large values indicate unstable
importance weights. We refer to
Refs.~\cite{vehtari2017practical,vehtari2024pareto} for the full
derivation.

We performed a PSIS-LOO comparison across the full kernel set using the full real-data \(z\leq2\) sample. The resulting stacking weights were sharply concentrated on a single kernel for each prior--method combination. We used this comparison to identify the dominant kernel in each case, and then repeated the reconstructions with those kernels under stricter convergence checks. The results reported in Sec.~\ref{sec:results} are based on these refined reconstructions. The full kernel-by-kernel PSIS-LOO results are listed in Table~\ref{tab:psis_table}.

\begin{table}[!t]
\centering
\small
\setlength{\tabcolsep}{2pt}
\caption{PSIS-LOO comparison across covariance kernels for the full \(z\leq2\) real-data sample. Method~A is shown for each external prior set; Method~B is listed once because its observed-\(H(z)\) PSIS-LOO quantities are identical across these three prior sets in this analysis.}
\label{tab:psis_table}
\begin{tabular}{@{}lcccc@{}}
\toprule
Case & Kernel & Weight & ELPD$_{\mathrm{LOO}}$ & SE \\
\midrule

\multirow{6}{*}{A: Planck2018}
& Cauchy             & 1 & -153.13 & 3.68 \\
& Mat\'ern $3/2$ & 0 & -153.18 & 3.66 \\
& Mat\'ern $7/2$ & 0 & -153.19 & 3.72 \\
& Mat\'ern $9/2$ & 0 & -153.25 & 3.71 \\
& RBF            & 0 & -153.25 & 3.74 \\
& Mat\'ern $5/2$ & 0 & -153.27 & 3.70 \\

\midrule

\multirow{6}{*}{A: DESI--BBN}
& Mat\'ern $5/2$ & 1 & -153.39 & 3.69 \\
& Mat\'ern $9/2$ & 0 & -153.40 & 3.72 \\
& Mat\'ern $3/2$ & 0 & -153.43 & 3.67 \\
& Mat\'ern $7/2$ & 0 & -153.47 & 3.72 \\
& Cauchy          & 0 & -153.47 & 3.70 \\
& RBF             & 0 & -153.59 & 3.74 \\

\midrule

\multirow{6}{*}{A: SH0ES--Pantheon}
& Cauchy             & 1 & -154.38 & 4.02 \\
& Mat\'ern $5/2$ & 0 & -154.43 & 4.04 \\
& Mat\'ern $3/2$ & 0 & -154.44 & 4.03 \\
& Mat\'ern $7/2$ & 0 & -154.57 & 4.04 \\
& Mat\'ern $9/2$ & 0 & -154.61 & 4.08 \\
& RBF            & 0 & -154.86 & 4.06 \\

\midrule

\multirow{6}{*}{B: Same for all priors}
& Mat\'ern $9/2$ & 1 & -153.48 & 3.96 \\
& Mat\'ern $7/2$ & 0 & -153.59 & 3.96 \\
& RBF             & 0 & -153.60 & 4.00 \\
& Mat\'ern $5/2$ & 0 & -153.63 & 3.96 \\
& Cauchy          & 0 & -153.66 & 3.97 \\
& Mat\'ern $3/2$ & 0 & -153.89 & 3.87 \\

\bottomrule
\end{tabular}
\end{table}

\section{Model comparison with log marginal likelihood}
\label{app:lml}

For completeness, Table~\ref{tab:lml_full} reports the log marginal
likelihood for the same full-sample kernel comparison. Across the
corresponding reconstructions, Method~A consistently yields larger LML values than
Method~B, corresponding to \(\Delta\log\mathcal Z\) of order \(8\) for representative best-kernel comparisons. Because marginal likelihoods
are more sensitive to prior specification than PSIS-LOO, we use them
only as a secondary diagnostic.

\begin{table}[!t]
\centering
\caption{Log marginal likelihood (LML) values for all reconstruction models. Method~B is listed once because its observed-\(H(z)\) likelihood is identical across the three external-prior sets; derived quantities remain prior dependent.}
\label{tab:lml_full}
\begin{tabular}{lllc}
\toprule
Prior & Method & Kernel & $\log p(\mathcal D|\mathcal M)$ \\
\midrule

\multirow{6}{*}{Planck2018}
& \multirow{6}{*}{A} & RBF                & -157.12 \\
&                       & Mat\'ern $3/2$   & -157.69 \\
&                       & Mat\'ern $5/2$   & -157.49 \\
&                       & Mat\'ern $7/2$   & -157.57 \\
&                       & Mat\'ern $9/2$   & -157.32 \\
&                       & Cauchy           & -157.60 \\
\midrule

\multirow{6}{*}{DESI--BBN}
& \multirow{6}{*}{A} & RBF                & -157.44 \\
&                       & Mat\'ern $3/2$   & -157.94 \\
&                       & Mat\'ern $5/2$   & -157.85 \\
&                       & Mat\'ern $7/2$   & -157.67 \\
&                       & Mat\'ern $9/2$   & -157.70 \\
&                       & Cauchy           & -157.97 \\
\midrule

\multirow{6}{*}{SH0ES--Pantheon}
& \multirow{6}{*}{A} & RBF                & -157.08 \\
&                       & Mat\'ern $3/2$   & -157.42 \\
&                       & Mat\'ern $5/2$   & -157.18 \\
&                       & Mat\'ern $7/2$   & -157.14 \\
&                       & Mat\'ern $9/2$   & -157.18 \\
&                       & Cauchy           & -157.37 \\
\midrule

\multirow{6}{*}{Same for all priors}
& \multirow{6}{*}{B} & RBF                & -165.53 \\
&                       & Mat\'ern $3/2$   & -166.78 \\
&                       & Mat\'ern $5/2$   & -165.89 \\
&                       & Mat\'ern $7/2$   & -165.73 \\
&                       & Mat\'ern $9/2$   & -165.64 \\
&                       & Cauchy           & -166.38 \\
\bottomrule
\end{tabular}
\end{table}





\section{Prior-expanded discrepancy tables}
\label{app:discrepancy_tables}
For completeness, Table~\ref{tab:discrepancy_appendix_full} reports the
prior-expanded discrepancy summaries for both real-data and mock comparisons. We use the same definition of $A_I$ and ROPE threshold $\delta=3$ as in the main text.
\clearpage
{
\begin{longtable*}{@{}lllcccc@{}}
\caption{Prior-expanded discrepancy summaries for real-data and mock comparisons. We report \(A_I=(1/|I|)\int_I |\Delta f|\,dz\) with \(\Delta f=f_A-f_B\) and ROPE threshold \(\delta=3\). Entries in the \(A_I\) column are written as \(m^{+(q84-m)}_{-(m-q16)}\), with \(m=q50\). Mock rows are realization-aggregated within each prior. The columns \(\max|d|\) and \(\max pd\) report the maximum pointwise effect size and posterior direction probability over the corresponding interval. Here ``baseline'' denotes the mock setup using the observed OHD redshift positions and reported \(\sigma_H\) values, while ``denser high-\(z\)'' denotes the mock setup in which this baseline configuration is retained but the \(z\ge1.5\) tail is sampled more densely and assigned smaller \(\sigma_H\) values.}
\label{tab:discrepancy_appendix_full}\\
\toprule
Case & Prior & $I$ & $A_I$ q50$^{+}_{-}$ & $P(A_I<3)$ & $\max|d|$ & $\max pd$ \\
\midrule
\endfirsthead
\toprule
\multicolumn{7}{@{}l@{}}{\textit{Table~\thetable\ (continued)}}\\
\midrule
Case & Prior & $I$ & $A_I$ q50$^{+}_{-}$ & $P(A_I<3)$ & $\max|d|$ & $\max pd$ \\
\midrule
\endhead
\midrule
\multicolumn{7}{r@{}}{\textit{Continued on next page}}\\
\endfoot
\bottomrule
\endlastfoot
Real: A vs B & Planck2018 & [0,1] & $1.99^{+1.93}_{-0.97}$ & 0.74 & 0.53 & 0.60 \\
 &  & [1,1.5] & $2.90^{+3.06}_{-1.67}$ & 0.52 & 0.41 & 0.57 \\
 &  & [1.5,2] & $4.37^{+3.48}_{-2.29}$ & 0.29 & 0.74 & 0.77 \\
 &  & [0,2] & $3.08^{+2.13}_{-1.30}$ & 0.48 & 0.74 & 0.77 \\
\cmidrule(lr){2-7}
 & DESI--BBN & [0,1] & $1.94^{+1.96}_{-0.95}$ & 0.74 & 0.52 & 0.59 \\
 &  & [1,1.5] & $2.84^{+3.01}_{-1.67}$ & 0.53 & 0.40 & 0.55 \\
 &  & [1.5,2] & $4.07^{+3.28}_{-2.12}$ & 0.33 & 0.67 & 0.75 \\
 &  & [0,2] & $2.99^{+1.96}_{-1.30}$ & 0.50 & 0.67 & 0.75 \\
\cmidrule(lr){2-7}
 & SH0ES--Pantheon & [0,1] & $2.06^{+1.82}_{-0.97}$ & 0.73 & 0.57 & 0.65 \\
 &  & [1,1.5] & $2.81^{+2.80}_{-1.52}$ & 0.54 & 0.29 & 0.59 \\
 &  & [1.5,2] & $4.54^{+3.27}_{-2.39}$ & 0.28 & 1.09 & 0.83 \\
 &  & [0,2] & $3.11^{+1.82}_{-1.23}$ & 0.47 & 1.09 & 0.83 \\
\midrule
Real: A-log-\(f\) vs A & Planck2018 & [0,1] & $0.87^{+1.31}_{-0.48}$ & 0.90 & 0.20 & 0.54 \\
 &  & [1,1.5] & $0.98^{+2.07}_{-0.61}$ & 0.84 & 0.18 & 0.52 \\
 &  & [1.5,2] & $1.02^{+1.63}_{-0.62}$ & 0.87 & 0.14 & 0.52 \\
 &  & [0,2] & $1.11^{+1.55}_{-0.60}$ & 0.86 & 0.20 & 0.54 \\
\cmidrule(lr){2-7}
 & DESI--BBN & [0,1] & $0.99^{+1.46}_{-0.55}$ & 0.89 & 0.32 & 0.55 \\
 &  & [1,1.5] & $1.21^{+2.33}_{-0.76}$ & 0.80 & 0.19 & 0.52 \\
 &  & [1.5,2] & $1.13^{+1.84}_{-0.68}$ & 0.84 & 0.18 & 0.54 \\
 &  & [0,2] & $1.27^{+1.68}_{-0.69}$ & 0.84 & 0.32 & 0.55 \\
\cmidrule(lr){2-7}
 & SH0ES--Pantheon & [0,1] & $1.92^{+5.67}_{-1.06}$ & 0.67 & 0.14 & 0.60 \\
 &  & [1,1.5] & $1.72^{+4.37}_{-1.03}$ & 0.70 & 0.11 & 0.56 \\
 &  & [1.5,2] & $1.68^{+3.00}_{-0.96}$ & 0.73 & 0.12 & 0.58 \\
 &  & [0,2] & $2.39^{+9.73}_{-1.27}$ & 0.61 & 0.14 & 0.60 \\
\midrule
Mock (baseline, fiducial \(\Lambda\)CDM) & Planck2018 & [0,1] & $0.36^{+0.21}_{-0.16}$ & 1.00 & 1.61 & 0.88 \\
 &  & [1,1.5] & $0.68^{+0.60}_{-0.35}$ & 1.00 & 0.59 & 0.67 \\
 &  & [1.5,2] & $2.19^{+1.62}_{-1.20}$ & 0.67 & 2.01 & 0.91 \\
 &  & [0,2] & $0.98^{+0.45}_{-0.39}$ & 1.00 & 2.08 & 0.94 \\
\cmidrule(lr){2-7}
 & DESI--BBN & [0,1] & $0.34^{+0.21}_{-0.15}$ & 1.00 & 1.41 & 0.85 \\
 &  & [1,1.5] & $0.66^{+0.63}_{-0.35}$ & 1.00 & 0.50 & 0.65 \\
 &  & [1.5,2] & $2.04^{+1.64}_{-1.12}$ & 0.73 & 1.77 & 0.89 \\
 &  & [0,2] & $0.92^{+0.44}_{-0.35}$ & 1.00 & 2.04 & 0.93 \\
\cmidrule(lr){2-7}
 & SH0ES--Pantheon & [0,1] & $0.44^{+0.21}_{-0.16}$ & 1.00 & 2.22 & 0.95 \\
 &  & [1,1.5] & $0.78^{+0.66}_{-0.39}$ & 1.00 & 0.93 & 0.73 \\
 &  & [1.5,2] & $2.95^{+1.85}_{-1.63}$ & 0.54 & 2.47 & 0.96 \\
 &  & [0,2] & $1.19^{+0.51}_{-0.45}$ & 1.00 & 2.65 & 0.97 \\
\midrule
Mock (baseline, mild \(w_0w_a\)) & Planck2018 & [0,1] & $0.33^{+0.20}_{-0.14}$ & 1.00 & 1.59 & 0.89 \\
 &  & [1,1.5] & $0.67^{+0.58}_{-0.36}$ & 1.00 & 0.57 & 0.66 \\
 &  & [1.5,2] & $2.25^{+1.59}_{-1.22}$ & 0.64 & 1.98 & 0.91 \\
 &  & [0,2] & $0.93^{+0.48}_{-0.36}$ & 1.00 & 2.03 & 0.92 \\
\cmidrule(lr){2-7}
 & DESI--BBN & [0,1] & $0.33^{+0.21}_{-0.14}$ & 1.00 & 1.52 & 0.87 \\
 &  & [1,1.5] & $0.66^{+0.59}_{-0.36}$ & 1.00 & 0.57 & 0.68 \\
 &  & [1.5,2] & $2.01^{+1.59}_{-1.09}$ & 0.70 & 1.78 & 0.89 \\
 &  & [0,2] & $0.88^{+0.47}_{-0.35}$ & 1.00 & 1.87 & 0.92 \\
\cmidrule(lr){2-7}
 & SH0ES--Pantheon & [0,1] & $0.41^{+0.21}_{-0.17}$ & 1.00 & 1.92 & 0.91 \\
 &  & [1,1.5] & $0.77^{+0.65}_{-0.41}$ & 1.00 & 0.89 & 0.73 \\
 &  & [1.5,2] & $2.93^{+1.86}_{-1.63}$ & 0.52 & 2.54 & 0.95 \\
 &  & [0,2] & $1.17^{+0.55}_{-0.42}$ & 1.00 & 2.56 & 0.96 \\
\midrule
Mock (denser high-\(z\), fiducial \(\Lambda\)CDM) & Planck2018 & [0,1] & $0.34^{+0.20}_{-0.15}$ & 1.00 & 1.40 & 0.85 \\
 &  & [1,1.5] & $0.54^{+0.51}_{-0.30}$ & 1.00 & 0.40 & 0.63 \\
 &  & [1.5,2] & $1.05^{+0.76}_{-0.53}$ & 0.98 & 1.29 & 0.85 \\
 &  & [0,2] & $0.62^{+0.28}_{-0.22}$ & 1.00 & 1.84 & 0.91 \\
\cmidrule(lr){2-7}
 & DESI--BBN & [0,1] & $0.34^{+0.21}_{-0.15}$ & 1.00 & 1.43 & 0.83 \\
 &  & [1,1.5] & $0.55^{+0.53}_{-0.30}$ & 1.00 & 0.45 & 0.63 \\
 &  & [1.5,2] & $0.97^{+0.83}_{-0.49}$ & 0.98 & 1.08 & 0.77 \\
 &  & [0,2] & $0.60^{+0.28}_{-0.21}$ & 1.00 & 1.60 & 0.87 \\
\cmidrule(lr){2-7}
 & SH0ES--Pantheon & [0,1] & $0.43^{+0.19}_{-0.17}$ & 1.00 & 2.20 & 0.95 \\
 &  & [1,1.5] & $0.65^{+0.61}_{-0.34}$ & 1.00 & 0.41 & 0.62 \\
 &  & [1.5,2] & $1.44^{+1.13}_{-0.75}$ & 0.90 & 1.95 & 0.91 \\
 &  & [0,2] & $0.78^{+0.34}_{-0.26}$ & 1.00 & 2.27 & 0.95 \\
\midrule
Mock (denser high-\(z\), mild \(w_0w_a\)) & Planck2018 & [0,1] & $0.32^{+0.19}_{-0.14}$ & 1.00 & 1.59 & 0.88 \\
 &  & [1,1.5] & $0.52^{+0.51}_{-0.29}$ & 1.00 & 0.40 & 0.63 \\
 &  & [1.5,2] & $0.93^{+0.78}_{-0.46}$ & 0.98 & 1.03 & 0.79 \\
 &  & [0,2] & $0.58^{+0.26}_{-0.20}$ & 1.00 & 1.74 & 0.90 \\
\cmidrule(lr){2-7}
 & DESI--BBN & [0,1] & $0.31^{+0.19}_{-0.14}$ & 1.00 & 1.54 & 0.87 \\
 &  & [1,1.5] & $0.53^{+0.54}_{-0.30}$ & 1.00 & 0.49 & 0.64 \\
 &  & [1.5,2] & $0.94^{+0.75}_{-0.47}$ & 0.98 & 0.92 & 0.75 \\
 &  & [0,2] & $0.58^{+0.29}_{-0.21}$ & 1.00 & 1.62 & 0.88 \\
\cmidrule(lr){2-7}
 & SH0ES--Pantheon & [0,1] & $0.40^{+0.21}_{-0.16}$ & 1.00 & 1.93 & 0.92 \\
 &  & [1,1.5] & $0.63^{+0.62}_{-0.34}$ & 1.00 & 0.36 & 0.61 \\
 &  & [1.5,2] & $1.31^{+1.14}_{-0.70}$ & 0.91 & 1.70 & 0.90 \\
 &  & [0,2] & $0.75^{+0.33}_{-0.26}$ & 1.00 & 2.16 & 0.94 \\
\end{longtable*}
}

\section{Real-data and realization-aggregated mock \(Om(z)\) tables}
\label{app:mock_om_table}
\normalsize
In Table~\ref{tab:om_all_models}, \(\Delta Om_{0.5,1.0}=Om(0.5)-Om(1.0)\), while \(\Delta Om_{0.5,1.5}=Om(0.5)-Om(1.5)\). Under the definition used in this work, which allows for spatial curvature, $\Lambda$CDM corresponds to a redshift-independent $Om(z)$ equal to $\Omega_{m0}$, so both $\Delta Om$ quantities are expected to be close to zero.
\begin{table*}[!b]
\centering
\caption{Posterior medians and central $68\%$ credible intervals for the real-data $Om(z)$ diagnostic in the full $z\leq2$ reconstruction. The \(\Delta Om\) summaries are computed from draw-by-draw differences, rather than by subtracting the marginal summaries of \(Om(z)\).}

\label{tab:om_all_models}
\begin{tabular}{@{}lllccccc@{}}
\toprule
Case & Prior & Method & $Om(0.5)$ & $Om(1.0)$ & $Om(1.5)$ & $\Delta Om_{0.5,1.0}$ & $\Delta Om_{0.5,1.5}$ \\
\midrule
Real ($z\leq2$) & Planck2018 & A & $0.27^{+0.39}_{-0.41}$ & $0.31^{+0.14}_{-0.12}$ & $0.32^{+0.08}_{-0.06}$ & $-0.04^{+0.34}_{-0.39}$ & $-0.05^{+0.37}_{-0.42}$ \\
 &  & B & $0.46^{+0.77}_{-0.66}$ & $0.36^{+0.43}_{-0.25}$ & $0.32^{+0.22}_{-0.18}$ & $0.00^{+0.69}_{-0.42}$ & $0.08^{+0.78}_{-0.63}$ \\
 &  & A-log-\(f\) & $0.28^{+0.15}_{-0.15}$ & $0.31^{+0.12}_{-0.05}$ & $0.32^{+0.06}_{-0.03}$ & $-0.04^{+0.10}_{-0.17}$ & $-0.04^{+0.13}_{-0.18}$ \\
\cmidrule(lr){2-8}
 & DESI--BBN & A & $0.21^{+0.51}_{-0.51}$ & $0.32^{+0.20}_{-0.18}$ & $0.31^{+0.10}_{-0.09}$ & $-0.10^{+0.45}_{-0.52}$ & $-0.10^{+0.53}_{-0.58}$ \\
 &  & B & $0.44^{+0.77}_{-0.64}$ & $0.35^{+0.41}_{-0.24}$ & $0.31^{+0.22}_{-0.17}$ & $0.00^{+0.67}_{-0.41}$ & $0.07^{+0.76}_{-0.61}$ \\
 &  & A-log-\(f\) & $0.26^{+0.20}_{-0.17}$ & $0.30^{+0.19}_{-0.06}$ & $0.30^{+0.09}_{-0.03}$ & $-0.06^{+0.14}_{-0.22}$ & $-0.05^{+0.18}_{-0.20}$ \\
\cmidrule(lr){2-8}
 & SH0ES--Pantheon & A & $0.11^{+0.52}_{-0.53}$ & $0.28^{+0.21}_{-0.19}$ & $0.29^{+0.09}_{-0.11}$ & $-0.17^{+0.48}_{-0.52}$ & $-0.19^{+0.58}_{-0.52}$ \\
 &  & B & $0.32^{+0.68}_{-0.54}$ & $0.30^{+0.34}_{-0.22}$ & $0.26^{+0.19}_{-0.16}$ & $-0.03^{+0.58}_{-0.35}$ & $0.02^{+0.67}_{-0.53}$ \\
 &  & A-log-\(f\) & $0.09^{+0.30}_{-0.04}$ & $0.26^{+0.16}_{-0.02}$ & $0.30^{+0.09}_{-0.01}$ & $-0.17^{+0.22}_{-0.10}$ & $-0.21^{+0.27}_{-0.06}$ \\
\bottomrule
\end{tabular}
\end{table*}

For clarity, Table~\ref{tab:om_all_models} reports only real-data results. The realization-aggregated mock summary used for quantitative interpretation is given in Table~\ref{tab:mock_om_agg}.

\begin{table*}[p]
\centering
\renewcommand{\arraystretch}{1.0}
\caption{Mock-ensemble summary for the \(Om(z)\) diagnostic. 
For each truth model, sampling configuration, prior, and method, 
\(\widetilde{\Delta Om}_{ij}\) denotes the median across mock data sets, 
with the superscript and subscript giving the 16th--84th percentile spread. 
The coverage fraction \(C_{68}^{0.5,1.5}\) is the fraction of mock data sets 
for which the posterior 68\% interval contains the injected truth value of 
\(\Delta Om_{0.5,1.5}\). Detailed definitions of \(\widetilde{\Delta Om}_{ij}\), the associated 16th--84th percentile spread, and \(C_{68}^{0.5,1.5}\) are given in the text below. Here ``baseline'' denotes 
the observed-sampling mock setup, which uses the observed OHD redshift 
positions and reported \(\sigma_H\) values, while ``denser high-\(z\)'' 
denotes the setup that keeps the lower-redshift sampling unchanged but samples 
the \(z\ge1.5\) tail more densely and assigns it smaller \(\sigma_H\) values.}
\label{tab:mock_om_agg}
{
\begin{tabular}{lllccc}
\toprule
Case & Prior & Method & $\widetilde{\Delta Om}_{0.5,1.0}$ & $\widetilde{\Delta Om}_{0.5,1.5}$ & $C_{68}^{0.5,1.5}$ \\
\midrule
fiducial $\Lambda$CDM (baseline) & Planck2018 & A & $-0.06^{+0.04}_{-0.10}$ & $-0.08^{+0.05}_{-0.09}$ & 0.35 \\
 &  & B & $-0.02^{+0.04}_{-0.07}$ & $0.01^{+0.06}_{-0.08}$ & 0.75 \\
\cmidrule(lr){2-6}
 & DESI--BBN & A & $-0.06^{+0.04}_{-0.12}$ & $-0.09^{+0.05}_{-0.08}$ & 0.40 \\
 &  & B & $-0.02^{+0.04}_{-0.07}$ & $0.00^{+0.06}_{-0.08}$ & 0.80 \\
\cmidrule(lr){2-6}
 & SH0ES--Pantheon & A & $-0.16^{+0.07}_{-0.09}$ & $-0.20^{+0.08}_{-0.08}$ & 0.00 \\
 &  & B & $-0.06^{+0.04}_{-0.06}$ & $-0.05^{+0.05}_{-0.07}$ & 0.55 \\
\midrule
fiducial $\Lambda$CDM (denser high-\(z\)) & Planck2018 & A & $-0.06^{+0.03}_{-0.09}$ & $-0.08^{+0.05}_{-0.06}$ & 0.40 \\
 &  & B & $-0.01^{+0.03}_{-0.07}$ & $-0.02^{+0.04}_{-0.07}$ & 0.65 \\
\cmidrule(lr){2-6}
 & DESI--BBN & A & $-0.06^{+0.03}_{-0.10}$ & $-0.09^{+0.05}_{-0.07}$ & 0.40 \\
 &  & B & $-0.02^{+0.03}_{-0.07}$ & $-0.02^{+0.04}_{-0.07}$ & 0.70 \\
\cmidrule(lr){2-6}
 & SH0ES--Pantheon & A & $-0.16^{+0.06}_{-0.09}$ & $-0.20^{+0.08}_{-0.04}$ & 0.05 \\
 &  & B & $-0.05^{+0.02}_{-0.06}$ & $-0.07^{+0.03}_{-0.06}$ & 0.35 \\
\midrule
mild $w_0w_a$ (baseline) & Planck2018 & A & $-0.03^{+0.02}_{-0.05}$ & $-0.04^{+0.03}_{-0.06}$ & 0.30 \\
 &  & B & $0.00^{+0.03}_{-0.01}$ & $0.03^{+0.03}_{-0.07}$ & 0.90 \\
\cmidrule(lr){2-6}
 & DESI--BBN & A & $-0.04^{+0.02}_{-0.04}$ & $-0.05^{+0.02}_{-0.06}$ & 0.30 \\
 &  & B & $-0.01^{+0.03}_{-0.01}$ & $0.03^{+0.03}_{-0.07}$ & 0.95 \\
\cmidrule(lr){2-6}
 & SH0ES--Pantheon & A & $-0.14^{+0.02}_{-0.03}$ & $-0.18^{+0.03}_{-0.06}$ & 0.00 \\
 &  & B & $-0.04^{+0.02}_{-0.01}$ & $-0.02^{+0.02}_{-0.06}$ & 0.75 \\
\midrule
mild $w_0w_a$ (denser high-\(z\)) & Planck2018 & A & $-0.04^{+0.02}_{-0.04}$ & $-0.06^{+0.04}_{-0.05}$ & 0.15 \\
 &  & B & $0.00^{+0.02}_{-0.02}$ & $0.00^{+0.03}_{-0.06}$ & 0.75 \\
\cmidrule(lr){2-6}
 & DESI--BBN & A & $-0.04^{+0.02}_{-0.05}$ & $-0.06^{+0.04}_{-0.06}$ & 0.25 \\
 &  & B & $-0.01^{+0.02}_{-0.02}$ & $0.00^{+0.03}_{-0.06}$ & 0.75 \\
\cmidrule(lr){2-6}
 & SH0ES--Pantheon & A & $-0.14^{+0.02}_{-0.03}$ & $-0.17^{+0.04}_{-0.04}$ & 0.00 \\
 &  & B & $-0.04^{+0.01}_{-0.01}$ & $-0.05^{+0.03}_{-0.06}$ & 0.35 \\

\bottomrule
\end{tabular}
}
\end{table*}
For the mock ensembles, let \(r\) label one independently noise-generated mock data set for a fixed truth model, sampling configuration, prior set, and reconstruction method. For each mock data set \(r\), we define
\begin{align}
\Delta Om_{0.5,1.0}^{(r)} &= Om^{(r)}(0.5)-Om^{(r)}(1.0), \\
\Delta Om_{0.5,1.5}^{(r)} &= Om^{(r)}(0.5)-Om^{(r)}(1.5).
\end{align}
The table reports the median and percentile band across mock data sets,
\begin{align}
\widetilde{\Delta Om}_{ij} &= \mathrm{median}_{r}\!\left[\Delta Om_{ij}^{(r)}\right], \\
\Delta Om_{ij}^{16\text{--}84} &= \mathrm{percentile}_{r}\!\left(\Delta Om_{ij}^{(r)};\,16,84\right),
\end{align}
and the \(68\%\)-coverage fraction
\begin{equation}
C_{68}^{0.5,1.5}
=
\frac{1}{N_{\rm mock}}
\sum_{r=1}^{N_{\rm mock}}
\mathbf{1}\!\left[
\ell_{68}^{(r)}
\le
\Delta Om_{0.5,1.5}^{\rm truth}
\le
u_{68}^{(r)}
\right],
\end{equation}
where \(N_{\rm mock}\) is the number of mock data sets, and \(\ell_{68}^{(r)}\) and \(u_{68}^{(r)}\) are the posterior \(68\%\) bounds obtained from mock data set \(r\). This coverage fraction is defined only for the mock ensembles, where the truth is known.

\section{Additional appendix material} 
\label{app:additional_figures}
For completeness, this section collects supplementary figures referenced in the main text.

\begin{figure}[!htbp]
\centering
\includegraphics[width=\columnwidth]{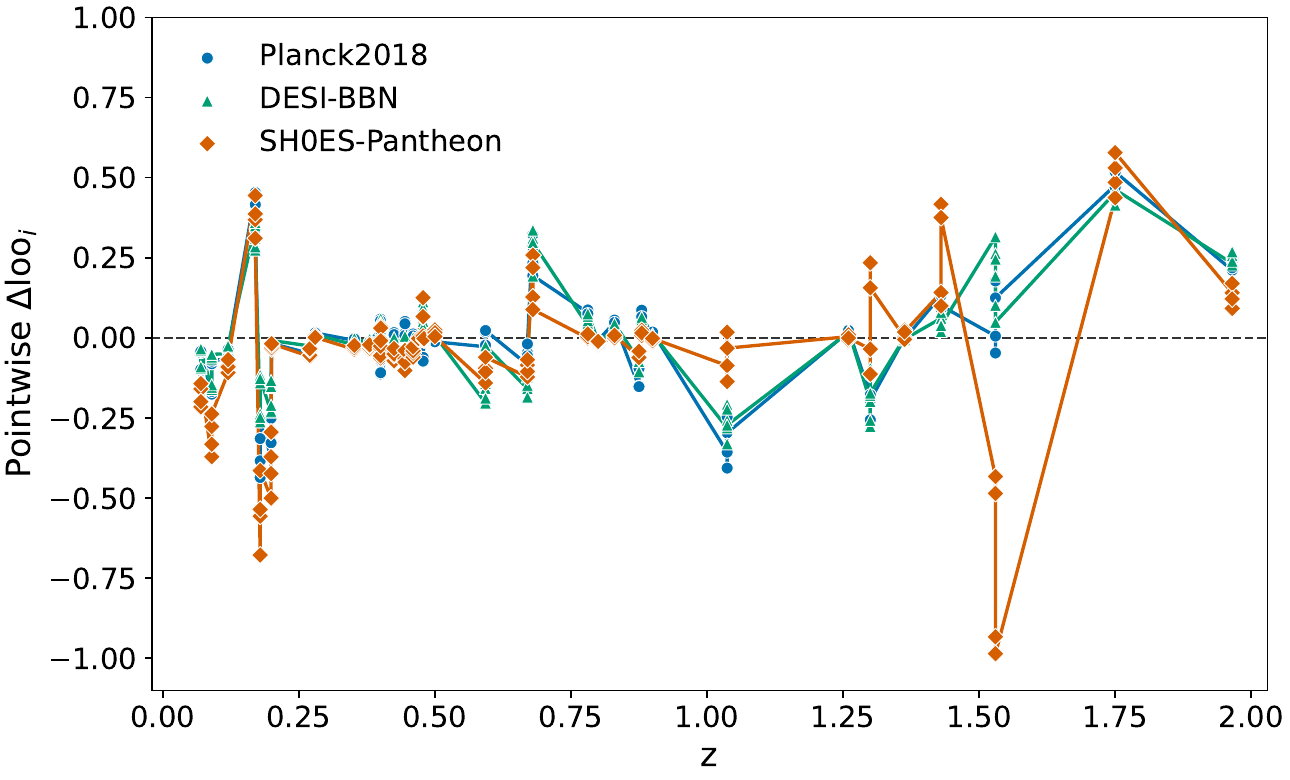}
\caption{
Pointwise PSIS-LOO contribution differences, \(\Delta\mathrm{loo}_i=\mathrm{loo}_{i,A}-\mathrm{loo}_{i,B}\), for the same-prior comparison between Methods~A and~B. For compactness, all three Method~A curves are compared against Method~B, which is the same for all priors in the observed \(H(z)\) predictive comparison.
}
\label{fig:delta_loo_pointwise}
\end{figure}

\begin{figure*}[p]
\centering
\subfloat[$z\leq 1$ truncated sample.]{%
\includegraphics[width=0.99\textwidth,height=0.40\textheight,keepaspectratio]{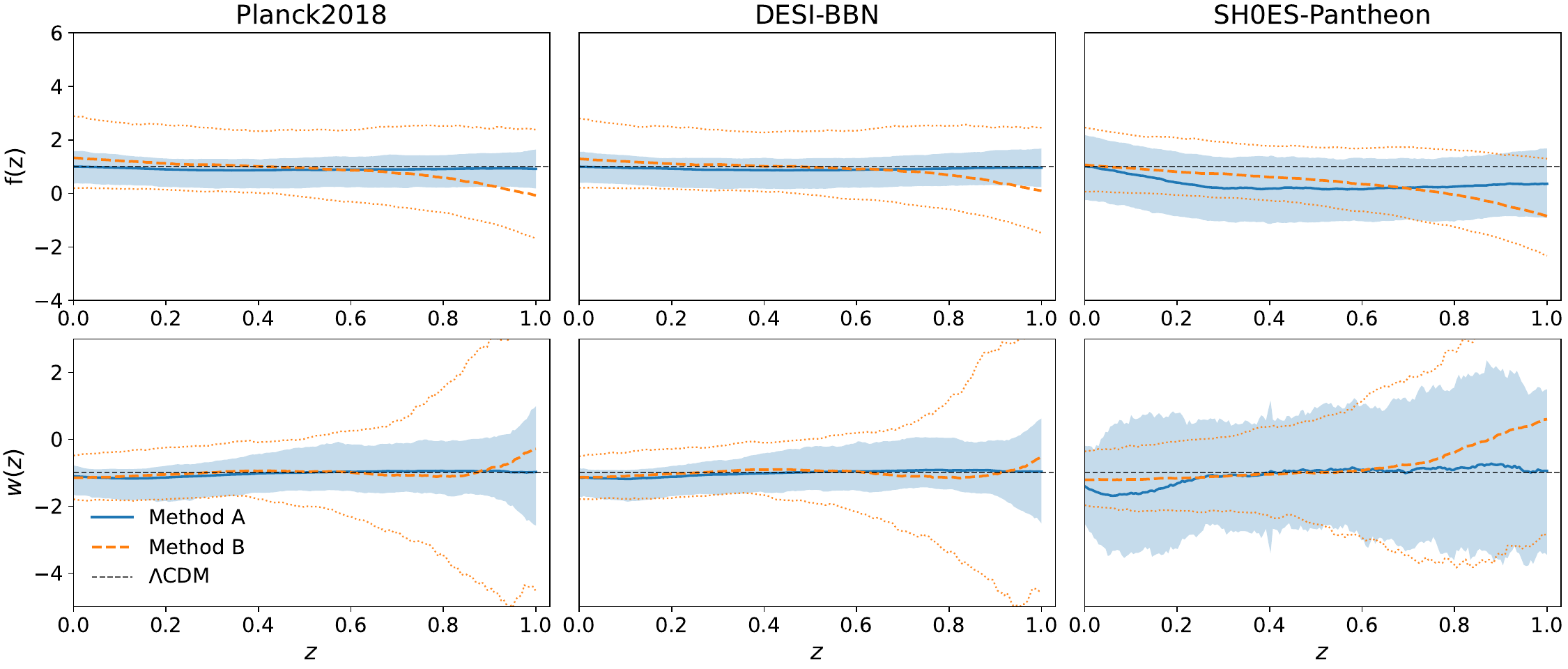}
\label{fig:fw_z1_appendix}
}\par\vspace{0.6em}
\subfloat[$z\leq 1.5$ truncated sample.]{%
\includegraphics[width=0.99\textwidth,height=0.40\textheight,keepaspectratio]{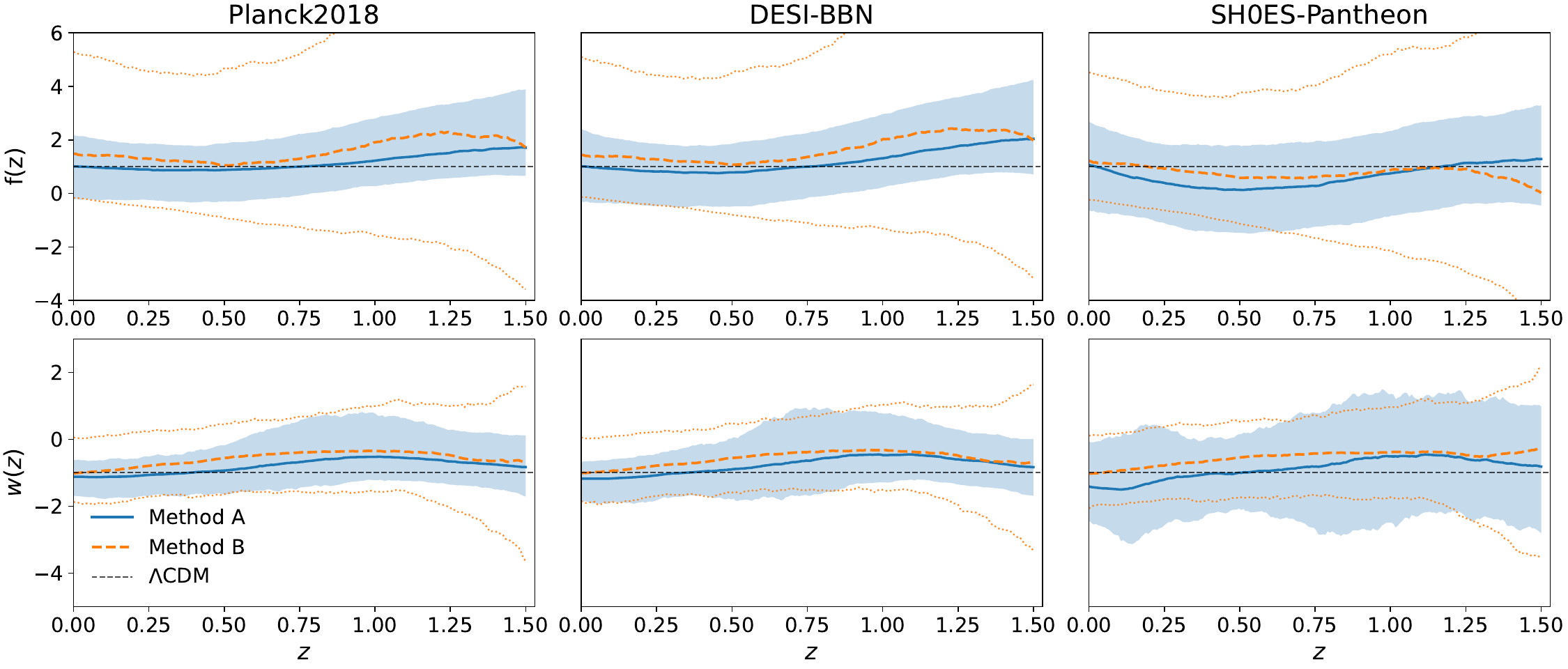}
\label{fig:fw_z15}
}
\caption{
Supplementary same-prior reconstructions between Methods~A and~B under nested redshift truncations.
Panels (a) and (b) correspond to $z\leq 1$ and $z\leq 1.5$, respectively.
Columns show the Planck2018, DESI--BBN, and SH0ES--Pantheon prior sets; the top row shows $f(z)$ and the bottom row shows $w(z)$.
Method~A is shown by blue solid curves with blue shaded \(68\%\) credible bands, while Method~B is shown by orange dashed curves with orange dashed \(68\%\) bounds.
The corresponding full-sample $z\leq 2$ result is shown in Fig.~\ref{fig:fw_main}.
}
\label{fig:fw_trunc_appendix}
\end{figure*}

\begin{figure*}[p]
\centering
\subfloat[$z\leq 1$.]{%
\includegraphics[width=0.99\textwidth,height=0.33\textheight,keepaspectratio]{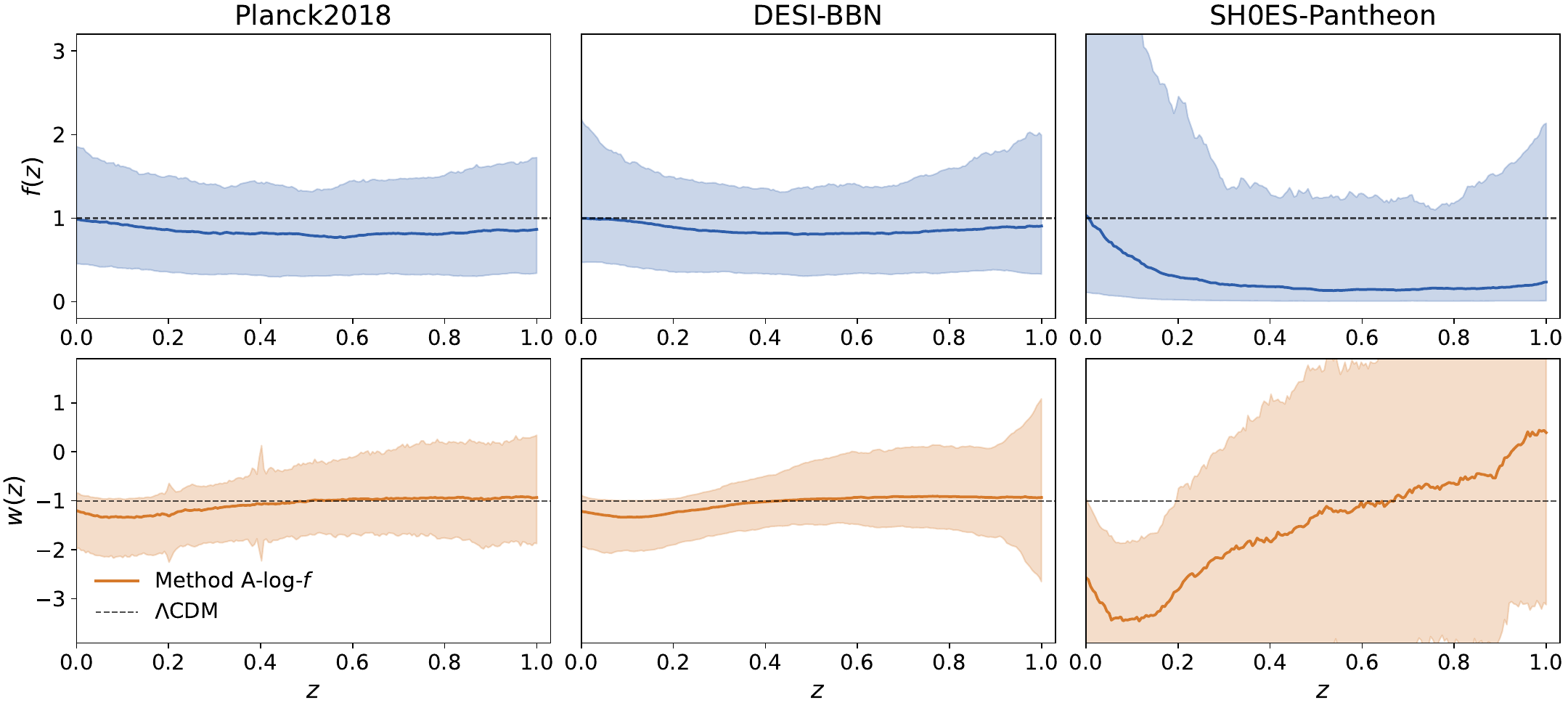}
\label{fig:logf_facet_appendix_z1}
}\\[0.2em]
\subfloat[$z\leq 1.5$.]{%
\includegraphics[width=0.99\textwidth,height=0.33\textheight,keepaspectratio]{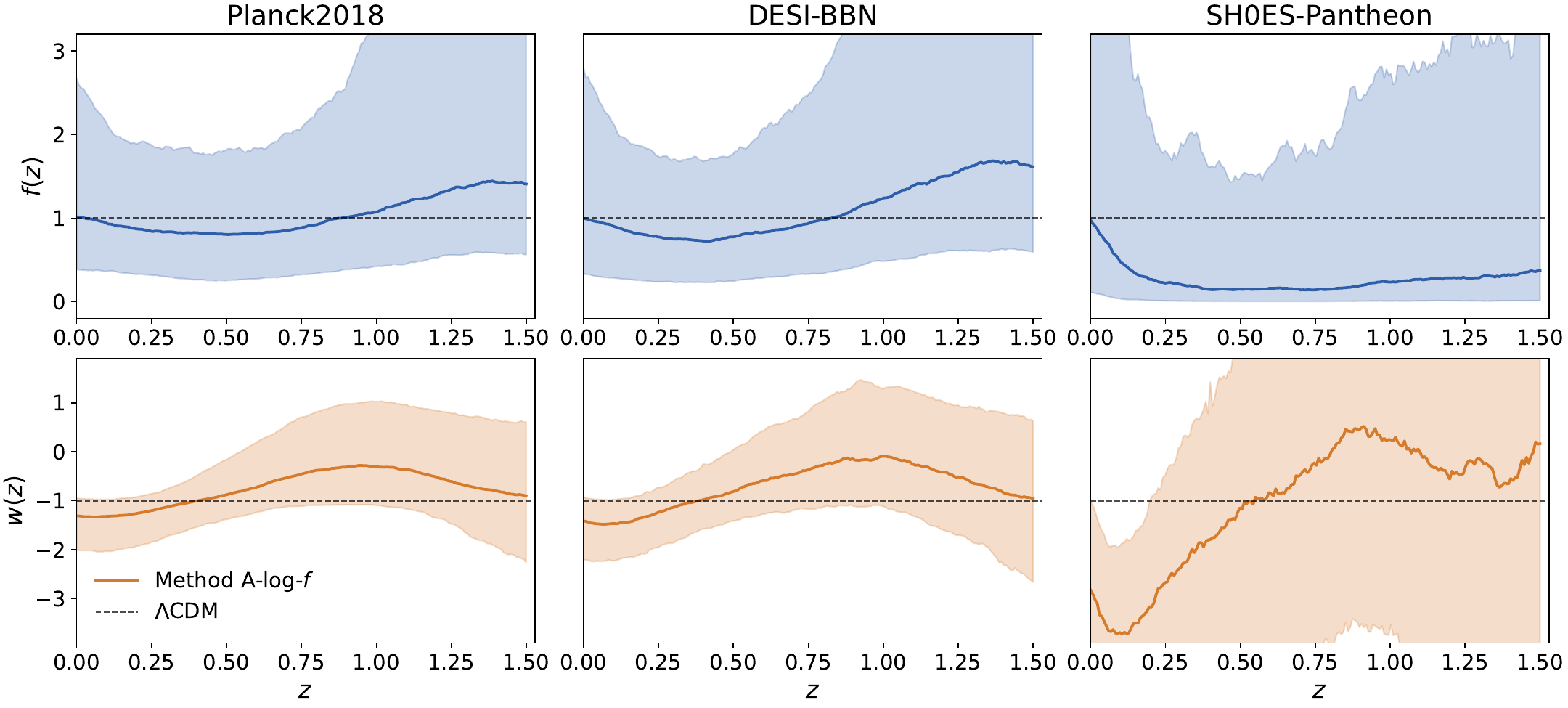}
\label{fig:logf_facet_appendix_z15}
}
\caption{
Supplementary A-log-\(f\) reconstructions for truncated samples.
Columns show the Planck2018, DESI--BBN, and SH0ES--Pantheon prior sets; the top row shows \(f(z)\), and the bottom row shows \(w(z)\).
The corresponding full-sample \(z\leq2\) A-log-\(f\) result is shown in Fig.~\ref{fig:logf_diagnostics} in the main text.
}
\label{fig:logf_facet_appendix}
\end{figure*}

\begin{figure*}[p]
\centering
\subfloat[Baseline, mild \(w_0w_a\).]{%
\includegraphics[width=0.9\textwidth,height=0.33\textheight,keepaspectratio]{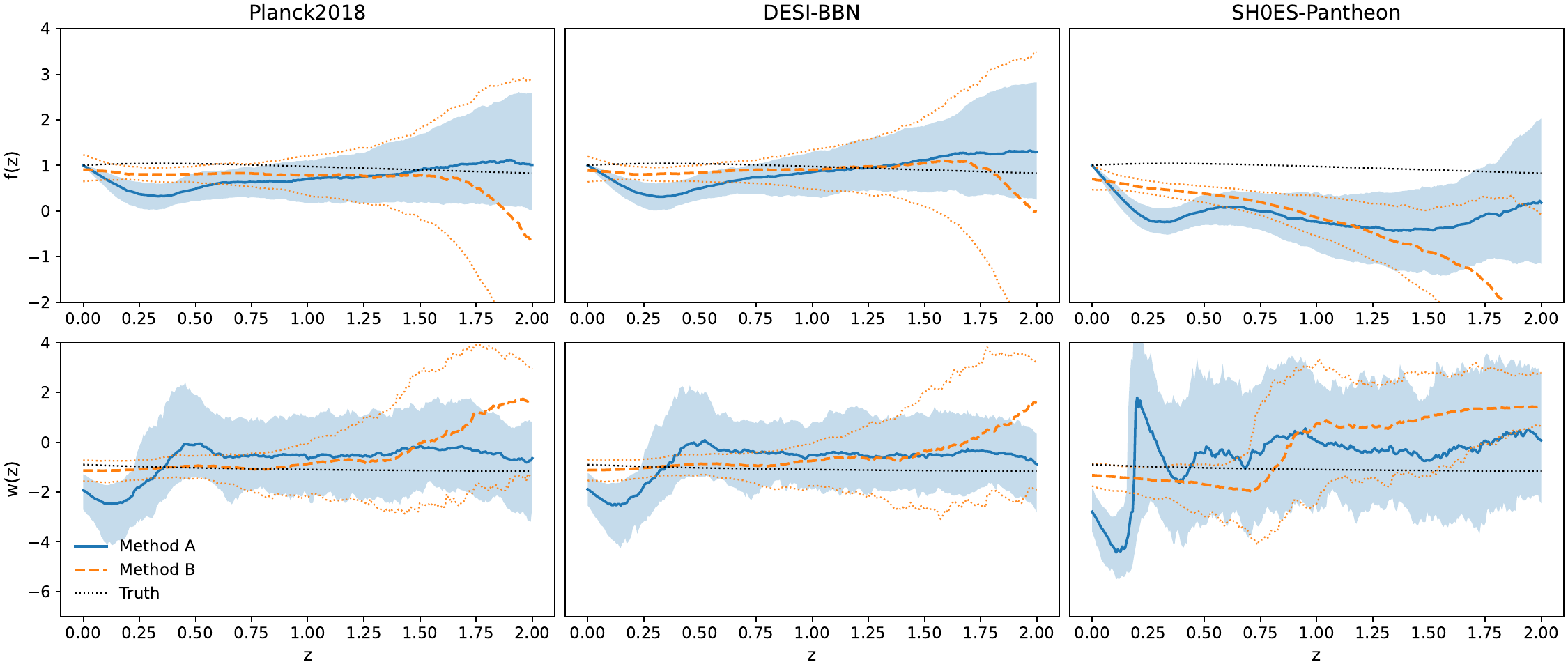}
\label{fig:mock_baseline_w0wa}
}\\[0.2em]
\subfloat[Denser high-\(z\), fiducial \(\Lambda\)CDM.]{%
\includegraphics[width=0.9\textwidth,height=0.33\textheight,keepaspectratio]{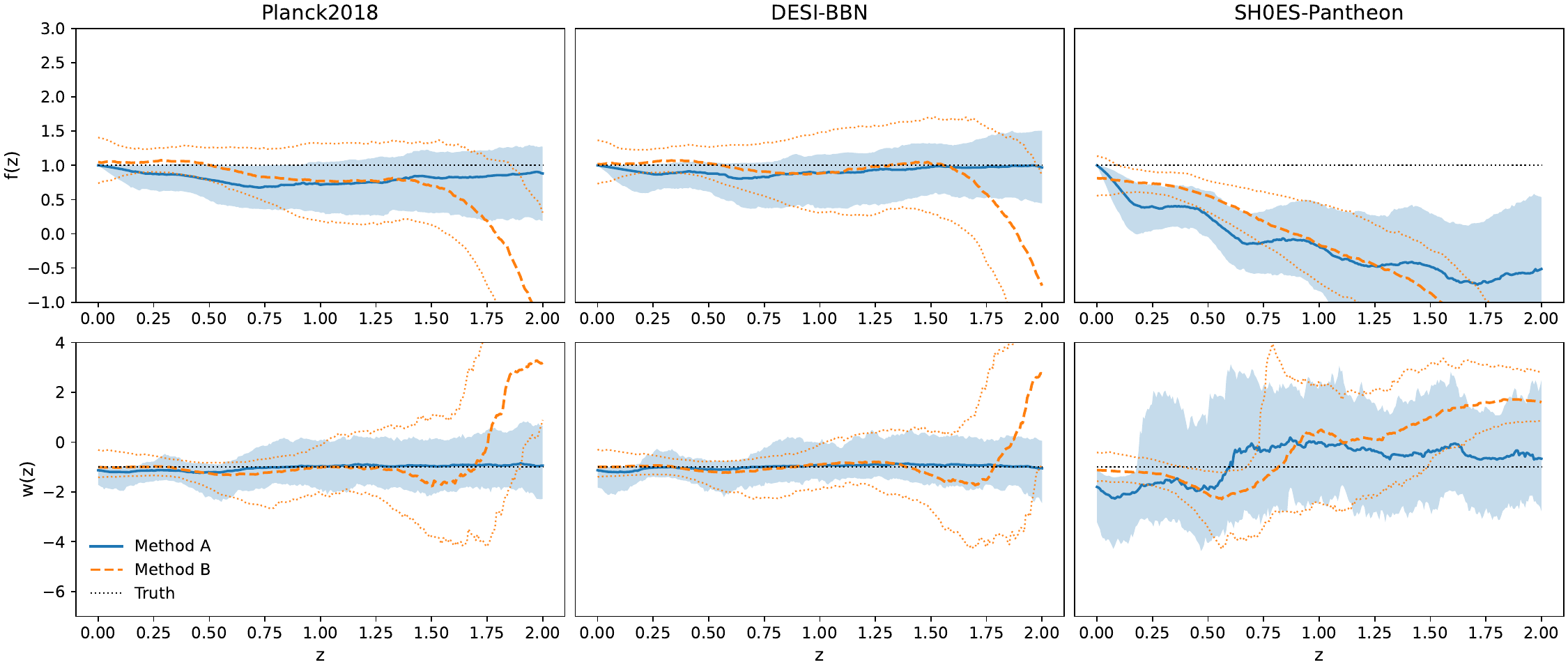}
\label{fig:mock_class3_lcdm}
}\\[0.2em]
\subfloat[Denser high-\(z\), mild \(w_0w_a\).]{%
\includegraphics[width=0.9\textwidth,height=0.33\textheight,keepaspectratio]{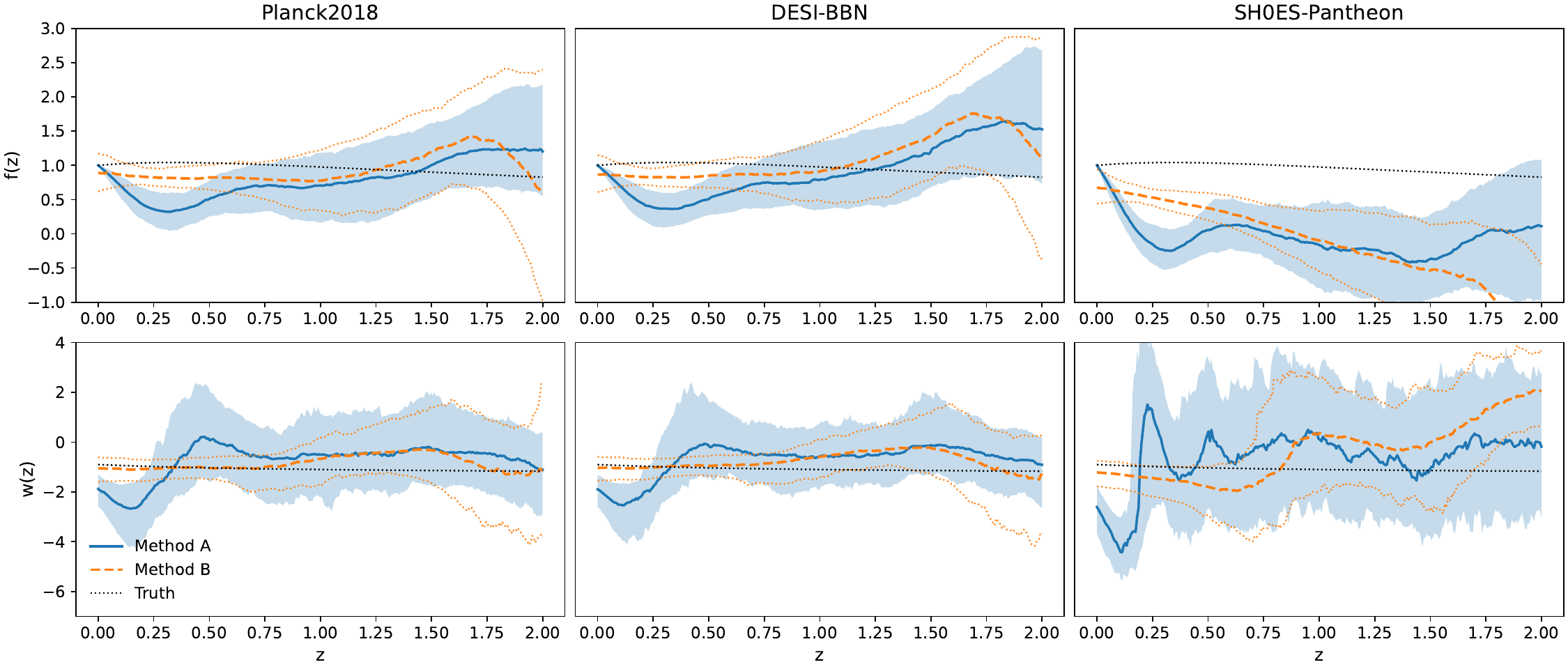}
\label{fig:mock_class3_w0wa}
}
\caption{
Supplementary mock-recovery examples completing the four-scenario set discussed in Sec.~\ref{sec:results_mock}: the observed-sampling mild \(w_0w_a\) case and the two denser high-\(z\) cases. In each panel the top row is \(f(z)\) and the bottom row is \(w(z)\). Curve styles and credible-band conventions are the same as in Fig.~\ref{fig:mock_baseline_lcdm}. Here ``baseline'' denotes the mock setup using the observed OHD redshift positions and reported \(\sigma_H\) values, while ``denser high-\(z\)'' denotes the setup in which this baseline configuration is retained but the \(z\ge1.5\) tail is sampled more densely and assigned smaller \(\sigma_H\) values.
}
\label{fig:mock_supp_examples}
\end{figure*}

\clearpage

\bibliographystyle{apsrev4-2}
\bibliography{main}
\end{document}